\documentclass{emulateapj}
\begin{document}

\newcommand{\MITRGB}{\rm M_{I}(TRGB)}
\newcommand{\MI}{\rm M_{I}}

\shorttitle{Star Formation History of M33}
\shortauthors{Barker, Sarajedini, Geisler, Harding \& Schommer}

\title{The Stellar Populations of M33's Outer Regions III: 
Star Formation History \footnotemark[1]} \footnotetext[1]{Based on observations made with the NASA/ESA Hubble Space Telescope, obtained at the Space Telescope Science Institute, which is operated by the Association of Universities for Research in Astronomy, Inc., under NASA contract NAS 5-26555. These observations are associated with program \# 9479.} 

\author{Michael K. Barker and Ata Sarajedini}
\affil{Department of Astronomy, University of Florida, Gainesville, FL 32611; mbarker@astro.ufl.edu, ata@astro.ufl.edu}

\author{Doug Geisler}
\affil{Grupo de Astronomia, Departamento de Fisica, Universidad de
Concepci\'{o}n, Casilla 160-C, Concepci\'{o}n, Chile; dgeisler@astro-udec.cl}

\author{Paul Harding}
\affil{Astronomy Department, Case Western Reserve University, 10900
Euclid Avenue, Cleveland, OH 44106; harding@dropbear.case.edu}

\author{Robert Schommer \footnotemark[2]} \footnotetext[2]{deceased.}
\affil{Cerro Tololo Inter-American Observatory, National Optical Astronomy
Observatories, Casilla 603, La Serena, Chile}

\begin{abstract} 

We present a detailed analysis of the 
star formation history (SFH) of three fields in M33 located
$\sim 4 - 6$ visual scale lengths from its nucleus.
These fields were imaged with the Advanced Camera for Surveys
on the {\it Hubble Space Telescope} and reach $\sim 2.5$ magnitudes 
below the red clump of core helium burning stars.
The observed color-magnitude diagrams are modeled as 
linear combinations of individual synthetic populations 
with different ages and metallicities.
To gain a better understanding of the systematic errors we
have conducted the analysis with two different sets of stellar
evolutionary tracks which we designate as 
Padova (Girardi et al.\ 2000) 
and Teramo (Pietrinferni et al.\ 2004).
The precise details of the results 
depend on which tracks are used but we can make several
conclusions that are fairly robust despite the differences.
Both sets of tracks predict the mean age to increase 
and the mean metallicity to decrease with radius.
Allowing age and metallicity to be free 
parameters and assuming star formation began 
$\sim 14$ Gyr ago, we find that the
mean age of all stars and stellar remnants 
increases from $\sim 6$ Gyr to $\sim 8$ Gyr and
the mean global metallicity decreases 
from $\sim -0.7$ to $\sim -0.9$.  
The fraction of stars formed by 4.5 Gyr ago
increases from $\sim 65\%$ to $\sim 80\%$.
The mean star formation rate $80 - 800$ Myr ago decreases from
$\sim 30\%$ of the lifetime average to
just $\sim 5\%$.
The random errors on these estimates are 
$\sim 10\%$, 1.0 Gyr, and 0.1 dex.  By comparing the results of
the two sets of stellar tracks for the real data and for
test populations with known SFH we have estimated the 
systematic errors to be $15\%$, 1.0 Gyr, and 0.2 dex.
These do not include uncertainties in the bolometric 
corrections or variations in $\alpha$-element abundance which 
deserve future study.

\end{abstract}


\section{Introduction}

The stellar populations of a galaxy are a fossil record 
of its formation and evolution and the various physical
processes involved.  Since the work of 
Eggen et al.\ (1962), 
the ages, kinematics, and chemical
compositions of stellar populations in the Galaxy have
proved crucial to our understanding of its evolution
(see Freeman \& Bland-Hawthorne 2002 for a recent review).
By the same token, such vital statistics are reshaping our
view of nearby galaxies especially those in the 
Local Group (LG; e.g., Sarajedini et al.\ 2000; 
Harbeck et al.\ 2001; Ferguson et al.\ 2002; 
Brown et al.\ 2003; Lanfranchi \& Matteucci 2004; 
Cole et al.\ 2005).

A critical tool in this advancement is the stellar color-magnitude
diagram (CMD).  
By comparing the observed distribution of stars in a
CMD with the predictions of stellar evolutionary theory
one can estimate their ages and metallicities.
For star clusters, 
such a comparison is straightforward relative to the more
complex task of disentangling the ages and metallicities
of stars in the general field.  Nevertheless, the study of
field populations has progressed tremendously due in part to 
advances in CMD analysis techniques and, in particular,
the technique of synthetic CMD fitting.  
Central to this technique is the use of theoretical stellar
evolutionary tracks with which one can
generate a model CMD for any arbitrary star formation 
history (SFH).  The model CMD can 
then be compared to the observed CMD to see how closely they
match and, thus, how closely the model SFH matches the true SFH.

Various forms of this technique have been applied to 
galaxies throughout the LG revealing a variety of
SFHs (e.g., Tosi et al.\ 1991; Bertelli et al.\ 1992; 
Aparicio et al.\ 1997; Dohm-Palmer et al.\ 1997; 
Tolstoy et al.\ 1998; Gallart et al.\ 1999; 
Hernandez et al.\ 2000; Olsen 1999; 
Miller et al.\ 2001; Wyder 2003; Skillman et al.\ 2003; 
Harris \& Zaritsky 2004; Holtzman et al.\ 1999; 
Mart\'{i}nez-Delgado et al.\ 1999; Dolphin 2002).
Surprisingly, though, it has yet to be applied in the
refereed literature to M33, the
third most massive galaxy in the LG.  M33 is a late-type
spiral galaxy making it the only other known spiral in the
LG besides the Galaxy and M31.  As such, it is a valuable
laboratory for studying disk galaxy evolution.

The first modern analysis of M33's field stars was carried out
by Mould \& Kristian (1986; MK86).  They used VI photometry of 
a field $\sim 20\arcmin$ southeast along M33's minor 
axis corresponding to a deprojected radius of
$R_{dp} \sim10$ kpc.  Because this field was located 
outside the optical
radius of M33's disk, it was assumed that they would be sampling the
halo population.  By comparing the observed red giant branch (RGB) 
to empirical RGBs of Galactic globular clusters (GGCs), they estimated 
the RGB stars in their field to have a mean metallicity of 
$\rm [Fe/H] = -2.2 \pm 0.8$.  
They concluded that the halo field population of M33 contains stars as 
metal-poor as the most metal-poor GGCs.

Since MK86, most stellar metallicity estimates in M33 have
utilized RGB stars in a similar manner.
Stephens \& Frogel (2002) resolved the RGB of M33's nucleus
in the near-infrared and
measured a metallicity of $-0.26$.
Kim et al.\ (2002) measured the metallicity at several different
locations throughout the inner disk and found it to decrease linearly
with galactocentric radius from $-0.6$ to $-0.9$.
Brooks et al.\ (2004) and Davidge (2003) measured
metallicities of $-1.3$ and $-1.0$, respectively, in the
far outer regions of M33 possibly sampling the halo.
In Tiede et al.\ (2004; Paper I) 
we used ground-based photometry
reaching the horizontal branch (HB) to study the metallicity
and spatial distribution of stars in a field coincident with
that studied by MK86.  With more accurate photometry we
concluded that the RGB metallicity was actually $\sim -1.0$.
In addition, the metallicity gradient was consistent with that
found by Kim et al.\ implying that this region was dominated
by disk rather than halo stars.

Comparitively little is known about the {\it ages} of
M33's stellar populations.  The ages are important because
not only do they tell us about the temporal and spatial 
progression of star formation but they also could affect the 
metallicity estimates summarized above.
Implicit in those estimates is the 
assumption that the RGB stars have the same mean age 
as the GGCs (i.e.\ $\sim 12$ Gyr).  
This assumption is necessary because of the 
age-metallicity degeneracy
of the RGB which is the property that increasing the age
has a similar effect on the RGB color as increasing 
the metallicity.  
The true metallicities could be higher than the above
estimates by a few tenths of a dex depending on the
true age (Salaris \& Girardi 2005).
Sarajedini et al.\ (2000)
presented CMDs for 10 of M33's halo globular clusters
and the background disk populations.  
Surprisingly, 8 of the 10 clusters showed red HB 
morphologies, indicating that they are possibly 
significantly younger than 12 Gyr.
The disk CMDs revealed mixed populations
in a wide range of evolutionary states suggesting
star formation in the disk has occurred over a long timespan.  
Therefore, it is certainly possible that M33's 
halo and disk red giants
do not have the same age as the GGCs.

In addition to stellar ages and metallicities, the large 
scale spatial distribution of M33's field stars can also provide clues 
to the system's structure and evolution.  Rowe et al.\ (2005)
mapped the distribution of different types of stars
from young, unevolved main sequence (MS) stars to 
older asymptotic giant branch (AGB) and RGB
stars.  They found the youngest stars to be concentrated
in spiral features while the oldest stars were more
evenly distributed throughout the disk 
demonstrating the migration of stars from
their birth sites.  Rowe et al.\ also used narrow-band
photometry to map the AGB populations and found the carbon
star density profile to extend out to a deprojected radius
of $R_{dp} \sim 50\arcmin - 60\arcmin$ where it appeared to flatten.  
The M-star profile was 
qualitatively similar with an unambiguous flattening
at $R_{dp} \sim 45\arcmin$ which they attributed
to foreground stars although we note that the density continued
to decline out to $R_{dp} \sim 90\arcmin$ perhaps
indicating a more extended component.
Finally, they concluded that the ratio of C-stars 
to M-stars, which is a rough
tracer of metallicity (but see Cioni et al.\ 2005 for 
a discussion of age effects), flattens out at $R_{dp} \sim 12\arcmin$.
They point out that such a flattening is consistent
with viscous disk formation models which predict gas
in the outer disk to be well mixed due to radial gas flows.

Observations
taken with the Advanced Camera for Surveys (ACS) on the
Hubble Space Telescope (HST) were presented in 
Barker et al.\ (2006; Paper II).  These data
covered three colinear fields located at projected radii
of $\sim 20 - 30\arcmin$ southeast of M33's nucleus,
the innermost of which overlapped with the field studied
in Paper I.  The field names were designated A1, A2, and A3
in order of increasing galactocentric distance.  The CMDs 
contained stars with ages from $\sim 100$ Myr to 
several Gyr or more.
The metallicity gradient was consistent with that of 
the inner disk and the stellar surface density dropped off
exponentially leading to the conclusion that the disk
extends out to $R_{dp} \sim 52\arcmin$ or 13 kpc at a
distance of 867 kpc.  
In addition, the radial scale length increased with age in a manner
similar to the vertical scale height of several nearby
late-type spirals.
For details of the observations, photometric reduction,
and artificial star tests we refer the reader to Paper II.
In the present 
study we use the same data presented in Paper II to make
quantitative estimates of the SFHs using the synthetic CMD
method mentioned above.

This paper is organized as follows.  In \S \ref{sec:synth}
we describe our implementation of the synthetic-CMD fitting method.
The results of applying this method to the ACS data
are presented in \S \ref{sec:results}.  We discuss the results
and their implications in \S \ref{sec:disc} and \S \ref{sec:conc}.
Lastly, in the Appendix we test the accuracy of the method
and its robustness against errors in the input parameters.

In this paper, age means lookback time 
(i.e., time from the present) and
the global metallicity is 
[M/H] $\equiv log[Z/Z_{\sun}]$ where $Z_{\sun} = 0.019$.

\section{Method}
\label{sec:synth}

We construct the model CMD from a linear combination of 
basis populations each of which represents the
predicted photometric distribution of stars within a 
certain range of ages and metallicities.  Each basis 
population forms a synthetic CMD which we create in a Monte Carlo
fashion using version 4.1 of the IAC-STAR program 
(Aparicio \& Gallart 2004).  
We use 5 metallicity bins 0.3 dex wide over the interval
$-1.7 \le \rm [M/H] \le -0.2$ and 9 age bins of width 0.25 dex 
in the range log(age/yr) = 7.90 - 10.15 (79.4 Myr - 14.1 Gyr).
This choice of age and metallicity binning is similar to
what has been used in other studies with photometry of
comparable depth 
(Wyder 2001, 2003; Dolphin et al.\ 2003).  It is
a compromise between precision, accuracy, and 
computational efficiency.  Smaller bins could increase
accuracy at the cost of losing precision, 
increasing noise in the solution, and increasing
computational time (Olsen 1999; Dolphin 2002).
The age bins are spaced 
logarithmically because the inherent 
precision decreases with age.  This occurs 
for two reasons.  First, the photometric errors and
incompleteness rate increase with magnitude
and the main sequence turnoff (MSTO) gets fainter with age.
More importantly, the spacing between the isochrones 
decreases with age.  

The input parameters required to make the sythetic CMDs
are the distance, extinction, initial mass 
function (IMF), binary fraction (f), and minimum mass ratio for
binaries (q).  In principle, it is
possible to solve for all these parameters simultaneously.  However,
given the sample sizes and photometric depth of the present
study, we elected to hold some of the parameters fixed 
while varying others.  

We adopt the default IMF in IAC-STAR which is a broken
power law with exponent $x = -1.35$ for 
$0.1 \leq M/M_{\sun} \leq 0.5$,
$x = -2.2$ for $0.5 \leq M/M_{\sun} \leq 1.0$, and $x = -2.7$ for
$1.0 \leq M/M_{\sun} \leq 120$.  This form of the IMF
is virtually identical to that derived by 
Kroupa et al.\ (1993) with the only difference 
being their low-mass slope is $-1.3$.
The slope of the low mass
end only affects the normalization
of the SFH because these stars lie below our detection limit.
Since we are observing a small range of masses at any given age we
cannot usefully constrain the slope at higher masses.  
Gallart et al.\ (1999) summarize recent observational
evidence for f = 0.4 and q = 0.6 so we adopt those values
in the present study.

Rather than hold the distance and extinction constant
we solve for them simultaneously with the SFH.  This
amounts to shifting the model CMDs in different directions 
and accounts for zero-point errors
in the theoretical isochrones and bolometric corrections.
We vary the distance over the range $(m-M)_0 = 24.50 - 24.80$
in steps of 0.10 mag.
This range encompasses most of M33's distance
measurements in the literature (Galleti et al.\ 2004).  The
extinction is varied over the range $A_V = 0.10 - 0.25$ in 
steps of 0.05 mag.  This range includes the 
Schlegel, Finkbeiner, \& Davis (1998)
value of $A_V = 0.15$ and also allows for some extinction internal
to M33.  For the extinction law we adopt 
$A_I = 1.31\ E(V-I)$ and $E(V-I) = 0.06$ as in Paper I
(Cardelli et al.\ 1989; von Hippel \& Sarajedini 1998).

We would like to vary the physical ingredients 
going into the theoretical isochrones since they probably
represent the largest possible sources of error.  Such ingredients
include the mixing length, helium enrichment, nuclear 
reaction rates, and opacities.  However, it is not yet
computationally feasible to do this and significant degeneracies
exist between the ingredients that could hinder such attempts.
The closest we can get to performing such an experiment
is to fit the CMDs using different sets of isochrones
with different physical ingredients and to compare the
results.  To that end we made separate fits using the
Padova (Girardi et al.\ 2000) and Teramo 
(Pietrinferni et al. 2004) isochrones transformed to the
observational plane with the Castelli \& Kurucz (2003)
$UBVRIJHKL$ bolometric correction library.

\begin{figure*}
\epsscale{0.95}
\plotone{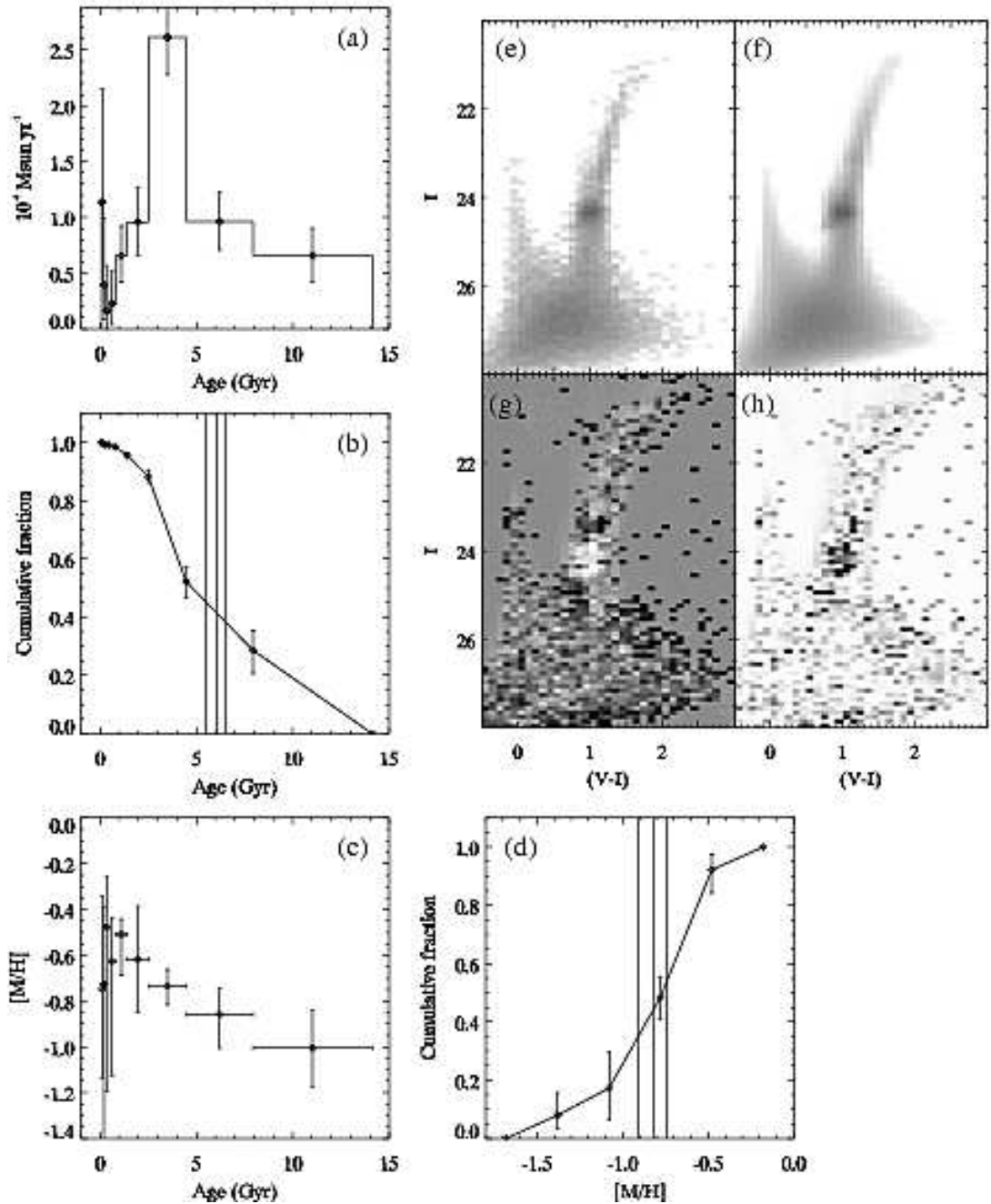}
\caption{SFH results for A1 using the Padova tracks
(see text for details). \label{fig:A1_padova_bad}}
\end{figure*}

\begin{figure*}
\epsscale{0.95}
\plotone{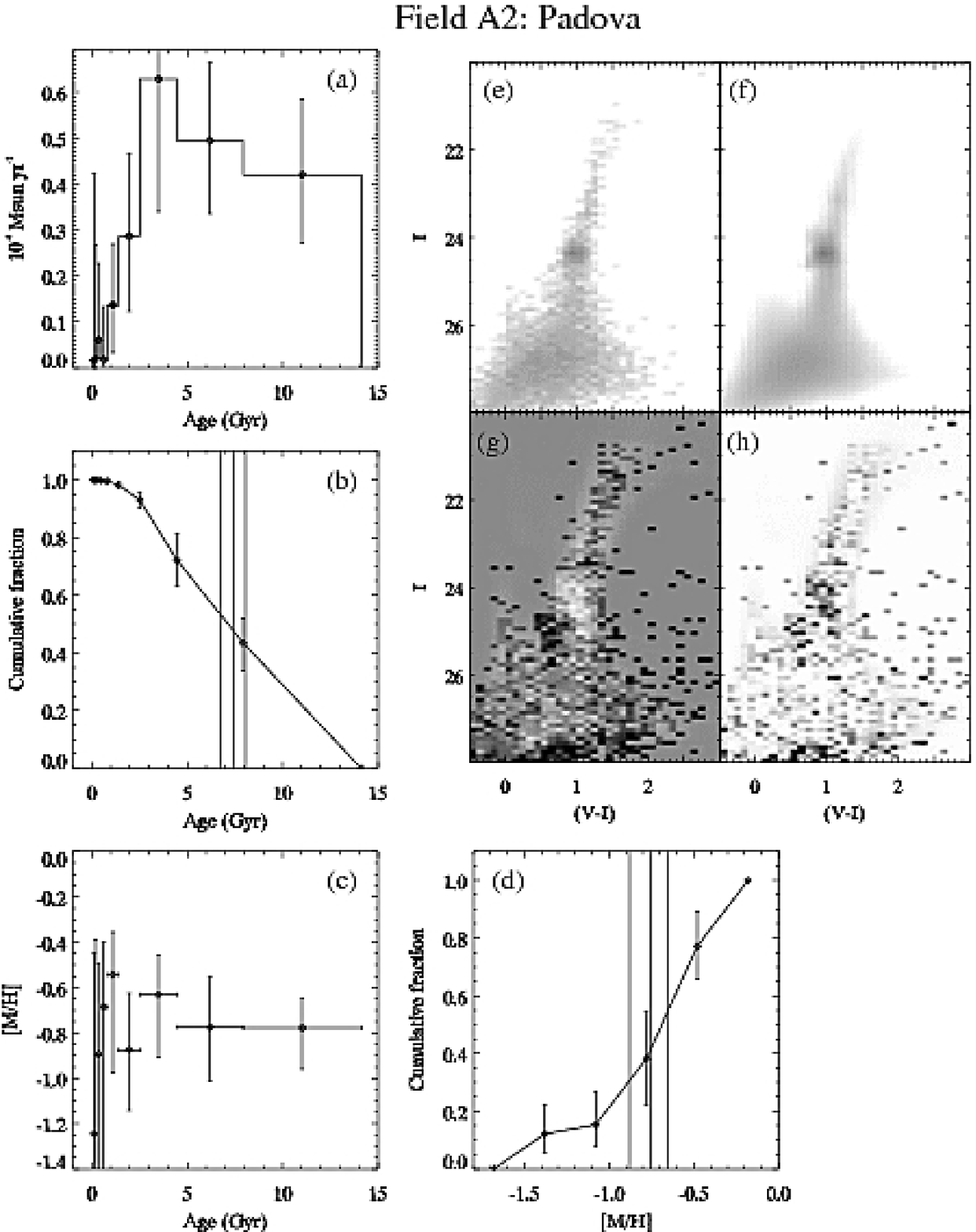}
\caption{Same as Fig.\ \ref{fig:A1_padova_bad} but for A2. \label{fig:A2_padova_bad}}
\end{figure*}

\begin{figure*}
\epsscale{0.95}
\plotone{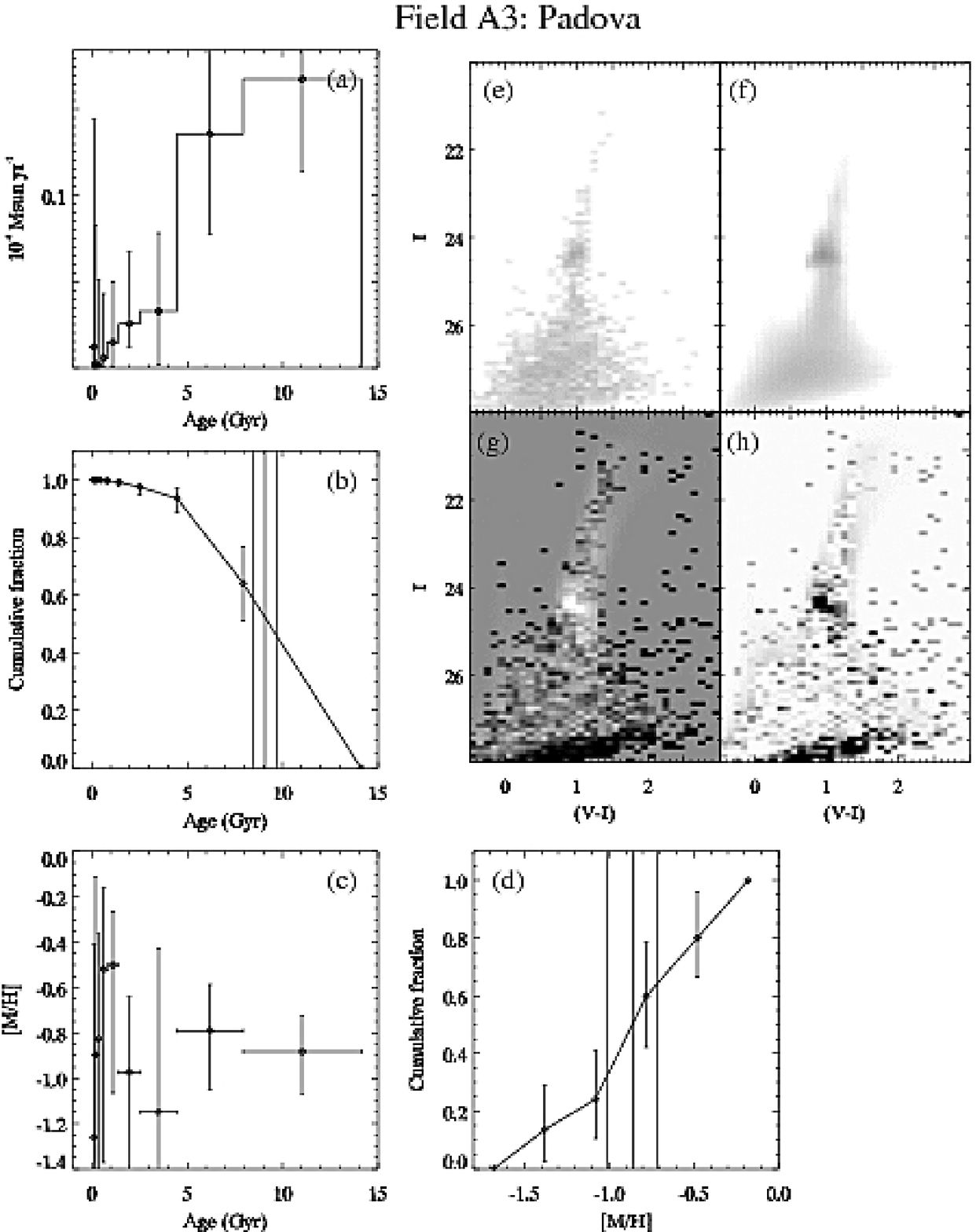}
\caption{Same as Fig.\ \ref{fig:A1_padova_bad} but for A3. \label{fig:A3_padova_bad}}
\end{figure*}

Ideally, we would work entirely in the native ACS/WFC filter
system.  IAC-STAR has an HST bolometric correction library
from Origlia \& Leitherer (2000) but it does not include
$F606W$ and, strictly speaking, it applies only to WFPC2.
As discussed in Sirianni et al.\ (2005) the WFPC2 and ACS
filter transmission curves are not identical.  Thus, we
are forced to work in the $UBVRI$ system using the synthetic
transformation of Sirianni et al.\ (2005).  In our experience
thus far, we have found
good agreement between HST and ground-based data in the $UBVRI$
system for several Galactic globular clusters with [Fe/H] $\lesssim -1.5$.  
Mackey, Payne \& Gilmore (2006)
found no significant systematic errors after transforming
to the ground-based system 
ACS/WFC photometry of two clusters in the 
Large Magellanic Cloud with metallicities
similar to the majority of our M33 stars ([Fe/H] $\sim -1.0$)
although it should be noted they used $F555W$ rather than $F606W$.

At higher metallicities the situation could be different.
Our investigations have found tentative evidence 
for a constant offset of $\sim -0.05$ mag in $V$ 
when comparing 47 Tuc ([Fe/H] $\approx -0.7$) 
ACS photometry to independent ground-based
data.
On the RGB, this offset translates into an uncertainty of
$\sim 0.1$ dex and $\sim 1.5$ Gyr at a metallicity of $-1.0$.
This offset is equal to the uncertainty of the photometric
zero-point of the Sirianni et al.\ synthetic transformation.  
Considering our coarse binning scheme in age, metallicity, and
in the CMD plane (as defined below) and given that a large fraction 
of our M33 stars have metallicities
less than 47 Tuc we believe that any such offset
is likely to have a small effect on our results especially
when compared to the effect of uncertainties 
in the theoretical stellar evolutionary tracks themselves.
We confirm this assertion in the Appendix where we 
fit a test population after manually inserting a 
V mag offset.

We employed StarFISH (Harris \& Zaritsky 2001)
to simulate the effects of observational errors in the synthetic
CMDs and to search for the best-fit model CMD.  
The artificial star tests described in Paper I allow us
to accurately quantify the photometric errors and completeness
rate as functions of both magnitude and color.
Each model star is
associated with a nearby artificial star.  If the artificial
star was recovered its magnitude shifts are assigned
to the model star otherwise the model star is discarded.  
Each synthetic CMD contains 
$\approx 1 \times 10^6$ stars before the simulation 
of observational errors.

The coefficients in the linear combination of synthetic CMDs
are proportional to the star formation rates (SFRs) at their respective ages
and metallicities.  
StarFISH uses a downhill simplex algorithm to solve for
the coefficients by minimizing a fitting statistic.  
We refer the reader to Harris \& Zaritsky (2001) for more 
details of the algorithm.  The model and data CMDs are divided
into square bins 0.1 mag on a side and the number
of model and data stars in each bin go into calculating
the fitting statistic.  This statistic is the
negative log-likelihood ratio for a Poisson distribution given by
$\Upsilon = 2\sum_{i}m_i - n_i + n_i {\rm ln}(n_i/m_i)$
where $m_i$ and $n_i$ are the number of model and data 
stars in CMD bin $i$, respectively.  The properties of this 
parameter have been discussed in various studies 
(e.g., Mighell 1999; Hauschild \& Jentschel 2001; 
Dolphin 2002).  For large $m_i$ it is well approximated by
the commonly used $\chi^2$.

Included in the linear combination of synthetic CMDs
is a bad-point CMD which we fit to
cosmic rays, hot pixels, foreground stars, 
and other objects in the observed CMD 
that cannot be reproduced by IAC-STAR 
(see also Dolphin 2002).  This has the form of a uniform
distribution that contributes $\sim 0.05$ ``stars'' to
each CMD bin resulting in $\sim 120$ over 
the entire CMD.  It also has the benefit
of preventing $\Upsilon$ from diverging when $m_i = 0$ and
$n_i > 0$.

Following Gallart et al.\ (1999) and Wyder (2001), the global best-fit 
model is the weighted average 
of all acceptable individual solutions, each of which
corresponds to a particular combination 
of distance and reddening.  A solution is acceptable if
it lies within $1\sigma$ of the best-fit.  
For each individual solution
StarFISH calculates the errors on the SFRs
by moving through the parameter space in many different 
directions until the fitting
statistic changes by $1\sigma$.  Thus, the errors
represent uncorrelated and correlated errors between the
amplitudes.  The errors of the individual best-fit solution are added
in quadrature with the spread of all acceptable solutions.  
In this way the errors of the global best-fit reflect 
correlations between age, metallicity, distance, and reddening.

\begin{figure*}
\epsscale{0.95}
\plotone{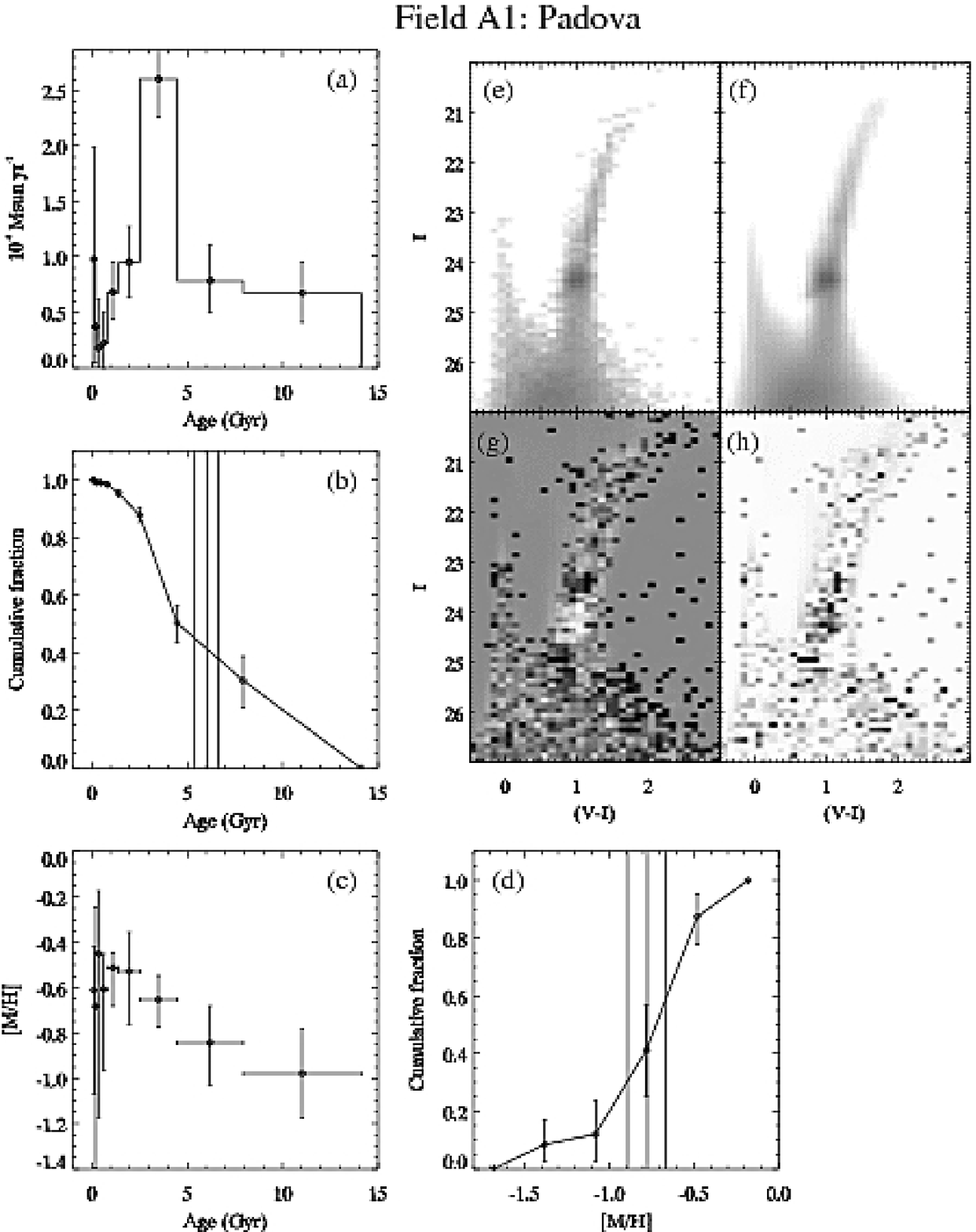}
\caption{SFH results for A1 using the Padova tracks and
after excluding the region $I > 27$ (see text for details). \label{fig:A1_padova}}
\end{figure*}

\begin{figure*}
\epsscale{0.95}
\plotone{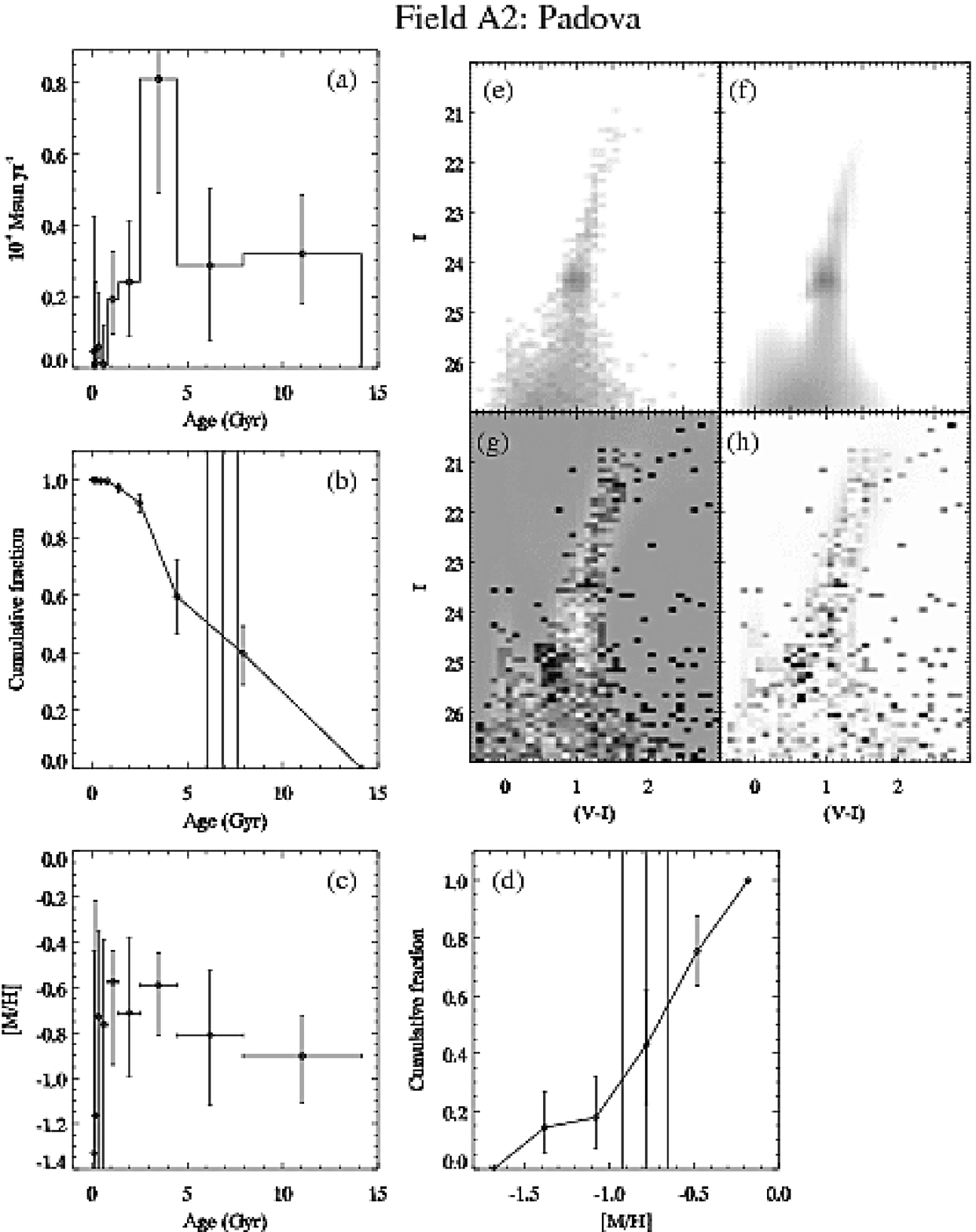}
\caption{Same as Fig.\ \ref{fig:A1_padova} but for A2. \label{fig:A2_padova}}
\end{figure*}

\begin{figure*}
\epsscale{0.95}
\plotone{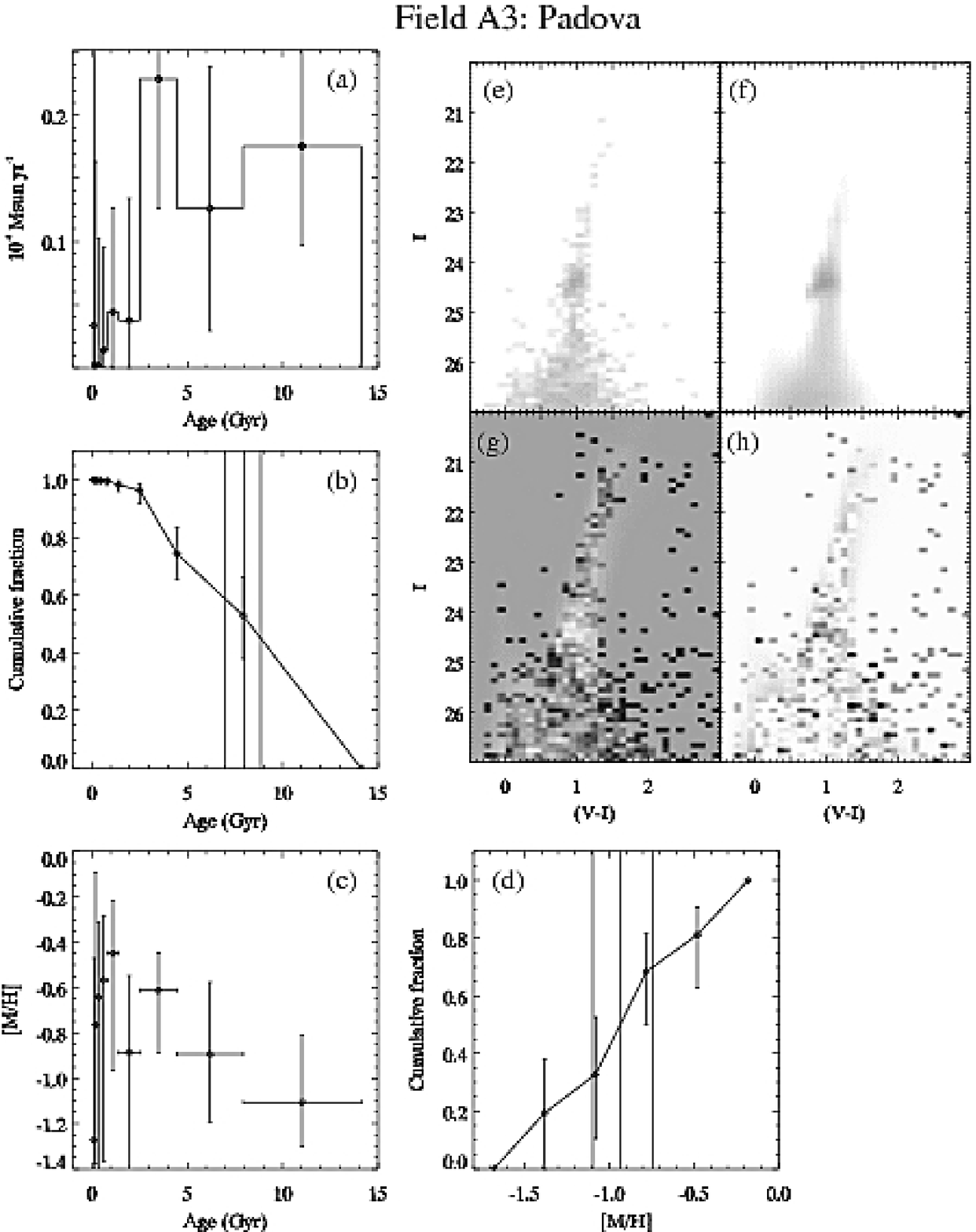}
\caption{Same as Fig.\ \ref{fig:A1_padova} but for A3. \label{fig:A3_padova}}
\end{figure*}

\begin{figure*}
\epsscale{1.0}
\plotone{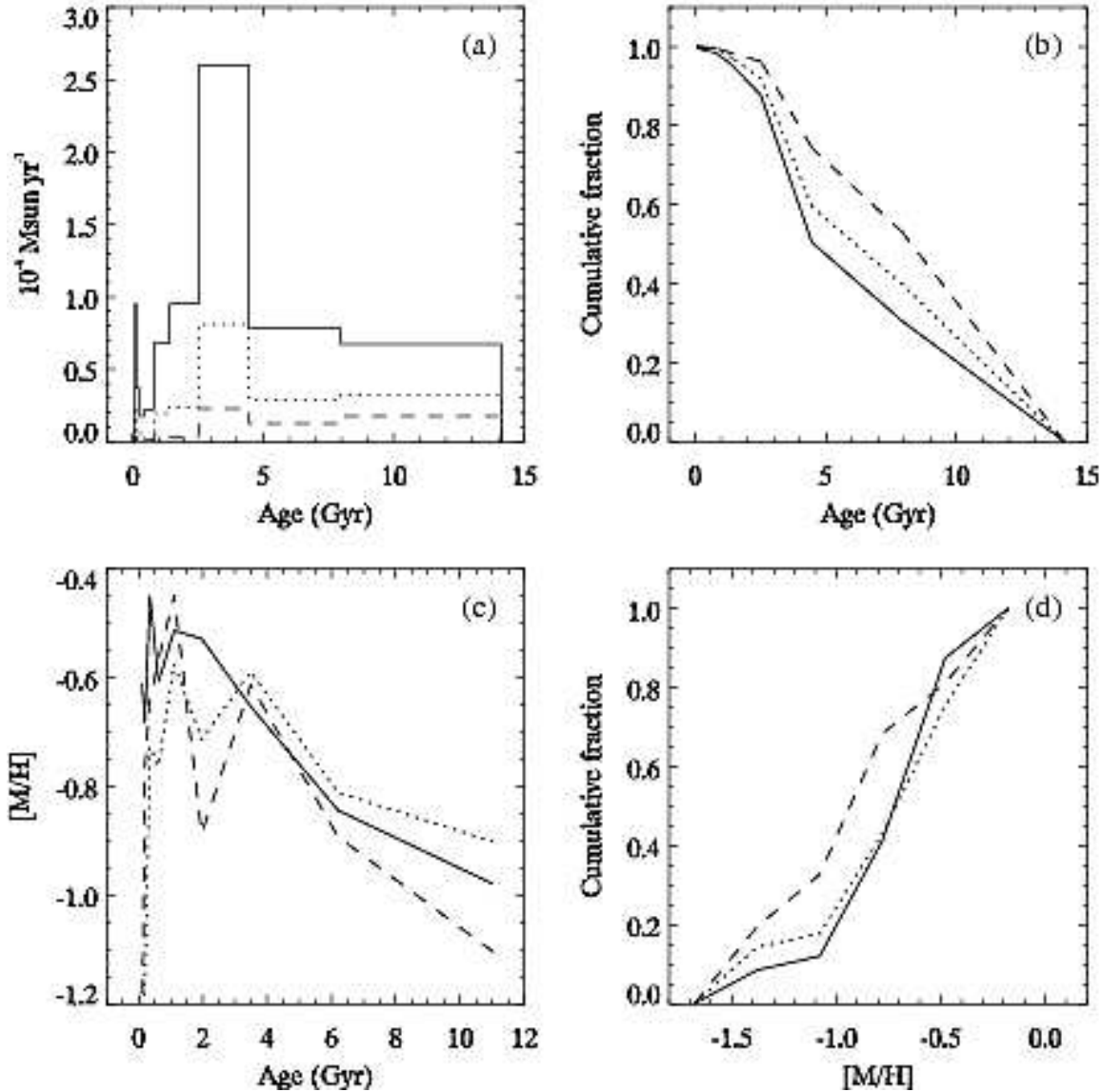}
\caption{Comparison of SFH
results for A1 (solid), A2 (dotted), and A3 (dashed). \label{fig:padova}}
\end{figure*}

\begin{figure*}
\epsscale{1.0}
\plotone{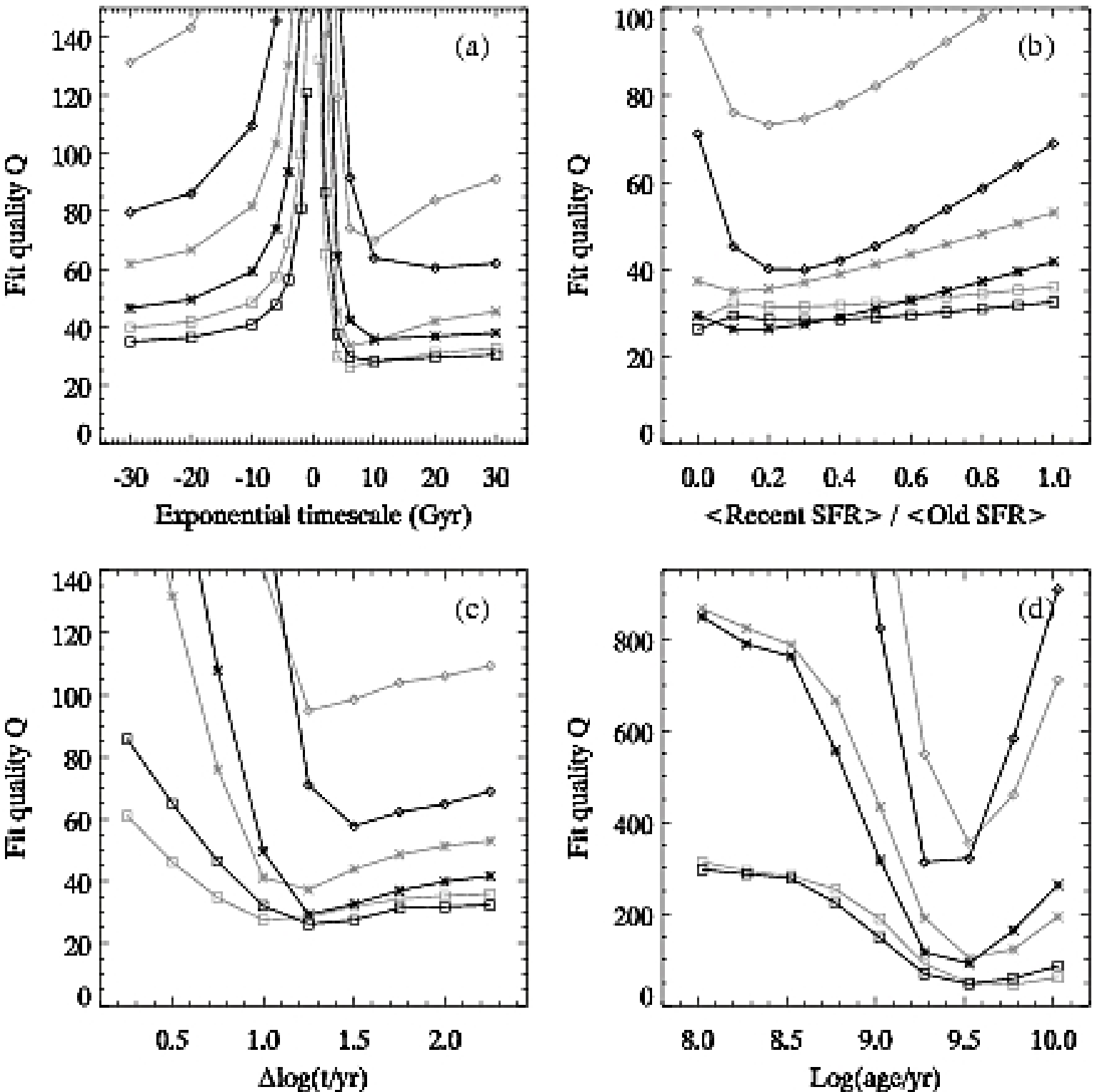}
\caption{Exploring parameter space by hand.  Each graph
shows how the fit quality varies with (a) exponential timescale,
(b) ratio of mean recent SFR to old SFR, (c) star formation
duration, and (d) age bin.  The point symbols are
diamonds, asterisks, and squares for fields A1, A2, and A3,
respectively.  Gray lines show the result of decreasing 
the metallicity (see text for details). \label{fig:padovaQ}}
\end{figure*}

The quality of the fit is measured by the parameter, $Q$,
which is the difference between $\Upsilon$ and its
expectation value in units of its standard deviation.
The expectation value depends on the model but is
approximately equal to the number of CMD bins contributing
to the fit minus the number of free parameters 
which include any nonzero SFH amplitudes (typically $\approx 25$)
plus distance and extinction.
The $Q$ parameter measures the likeliness of the data being
randomly drawn from the model (Dolphin 2002).
Only for comparison purposes we also calculate $\chi^2$ of
the best-fit defined by
$\chi^2 = \sum_{i}(n_i-m_i)^2/m_i$.
The reduced $\chi^2_{\nu} = \chi^2 / \nu$ where $\nu$ is
the number of significant CMD bins minus the number of free parameters.
Studies applying the synthetic CMD fitting method to real
stellar populations typically find values for $Q$ and $\chi^2_{\nu}$ 
in the range $\sim 1 - 5$ 
(Dolphin 2002; Harris \& Zaritsky 2004; 
Gallart et al.\ 1999; Skillman et al.\ 2003; 
Dolphin et al.\ 2003; Wyder 2001, 2003).

In the Appendix we demonstrate the effectiveness of the
method on several test populations.  In particular, we 
examine how errors in the
various input parameters affect the accuracy of the recovered SFH 
and how they contribute to systematic errors in the results.
Such tests are critical
to understanding the strengths and limitations of the method
and how to interpret the results.

\section{Results}
\label{sec:results}

\subsection{Padova tracks}
\label{sec:padova}

Figures $\ref{fig:A1_padova_bad} - \ref{fig:A3_padova_bad}$ 
present the results using 
the Padova tracks.  In each figure, panel (a)
shows the SFH and $1\sigma$ errors as explained before.
In panel (b) we show the age cumulative distribution 
function (age-CDF).
Panel (c) displays the age-metallicity relation (AMR) where
each point is the mean metallicity of all stars formed
in the corresponding age bin.  The horizontal errors
denote the age bin width.  
Panel (d) is the metallicity cumulative distribution
function (Z-CDF) of all stars ever formed.  
The vertical lines in panels (b) and (d) correspond
to the mean age and metallicity of all stars
and stellar remnants and the $1\sigma$ confidence intervals.
Also shown are the data CMD (e), model CMD (f), residuals (g), 
and significance (h).  The data and model CMDs are Hess diagrams 
(2-D histograms) on a logarithmic scale.
The residuals are on a scale where $-3\sigma$ is black 
and $+3\sigma$ is white and positive residuals mean 
the model is too high.

Upon inspection of the solutions, we see 
that the model CMDs underpredict the number
of stars fainter than $I = 27$.  
This indicates an approximately constant number 
of contaminants at $I > 27$ due to
non-stellar sources like unresolved background galaxies
and spurious noise artifacts.  
This contamination is
much less than the number of real stars in A1 but
becomes more significant in A2 and A3.  
The Teramo CMDs showed the
same discrepancy making it unlikely that the stellar
tracks are the cause.
As a test we repeated the entire photometric reduction
procedure and artificial star tests for A3 using more
stringent detection requirements.  
This allowed us to repeat the SFH analysis for A3 which
we found to yield better agreement between the
model and data CMDs but the model
still slightly underpredicted the number of stars
at the faint end.  
The resulting SFH was nearly
identical to that produced after excluding $I > 27$
in the original dataset
(see below) because there is little information
there anyway to constrain the solution.  
Since this region is also
beyond the $50\%$ completeness level, we will exclude
it for the remainder of the analysis.

\begin{figure*}
\epsscale{0.95}
\plotone{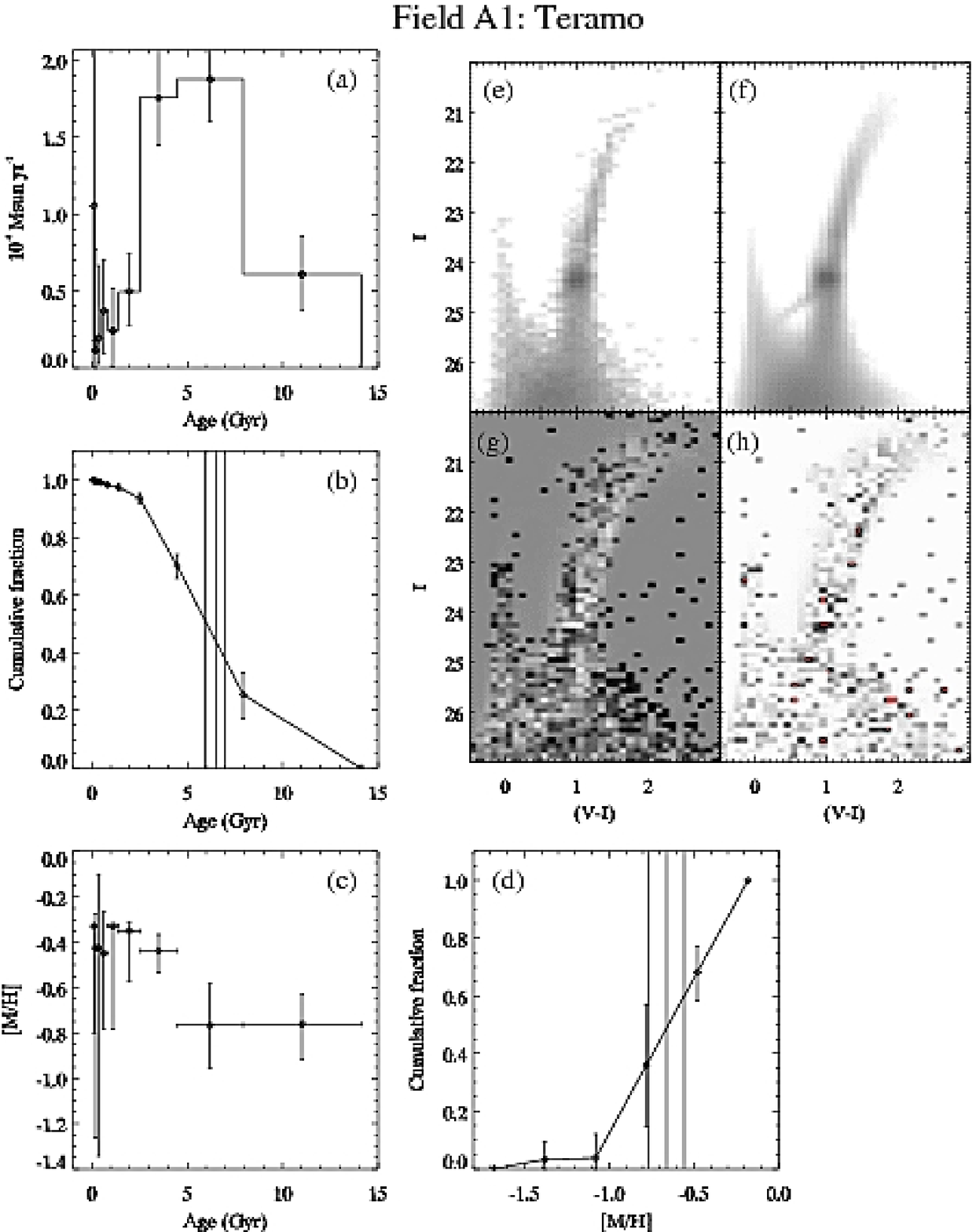}
\caption{SFH results for A1 using
the Teramo tracks (see text for details).}
\label{fig:A1_teramo}
\end{figure*}

\begin{figure*}
\epsscale{0.95}
\plotone{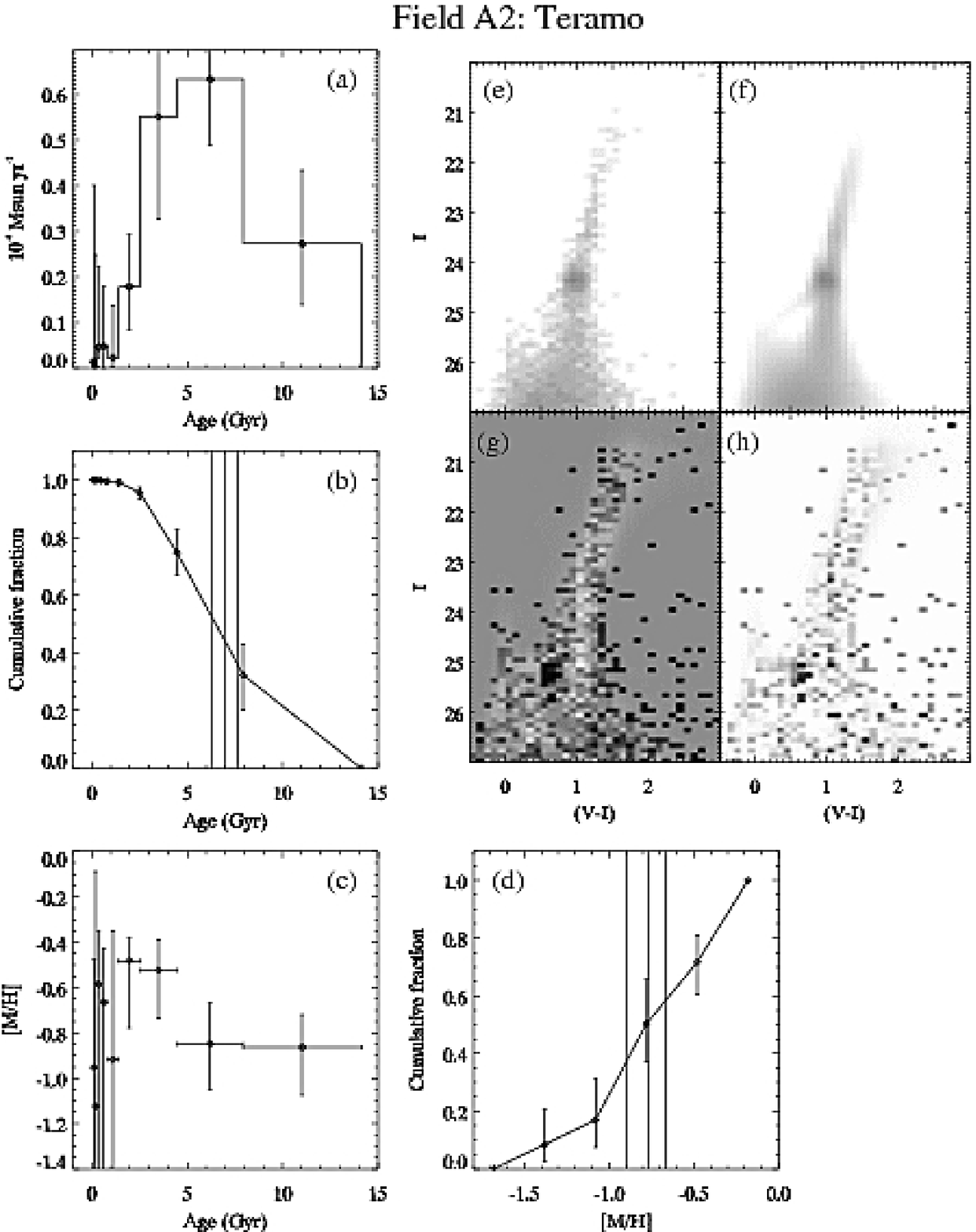}
\caption{Same as Fig.\ \ref{fig:A1_teramo} but for A2.}
\label{fig:A2_teramo}
\end{figure*}

\begin{figure*}
\epsscale{0.95}
\plotone{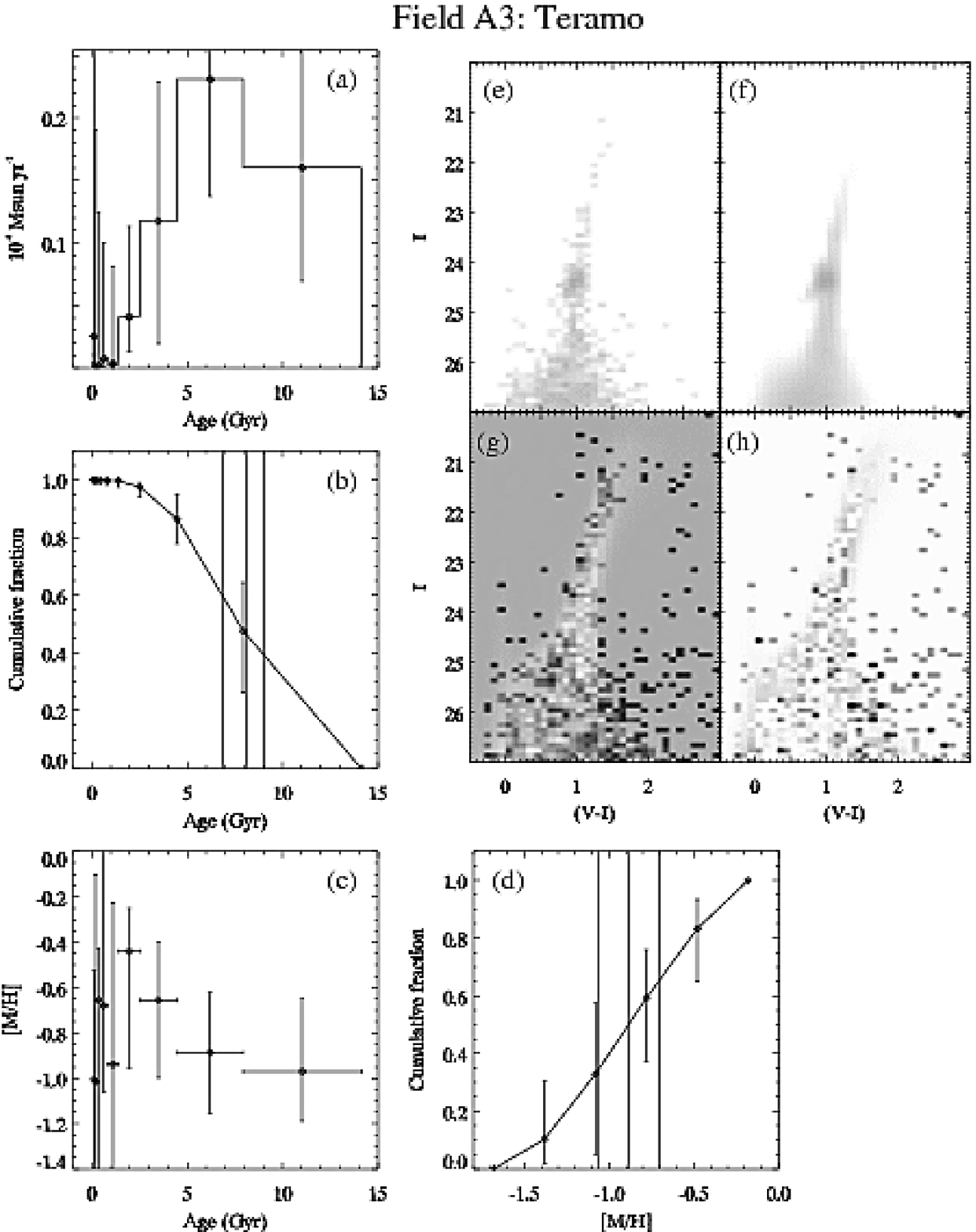}
\caption{Same as Fig.\ \ref{fig:A1_teramo} but for A3.}
\label{fig:A3_teramo}
\end{figure*}

\begin{figure*}
\epsscale{1.0}
\plotone{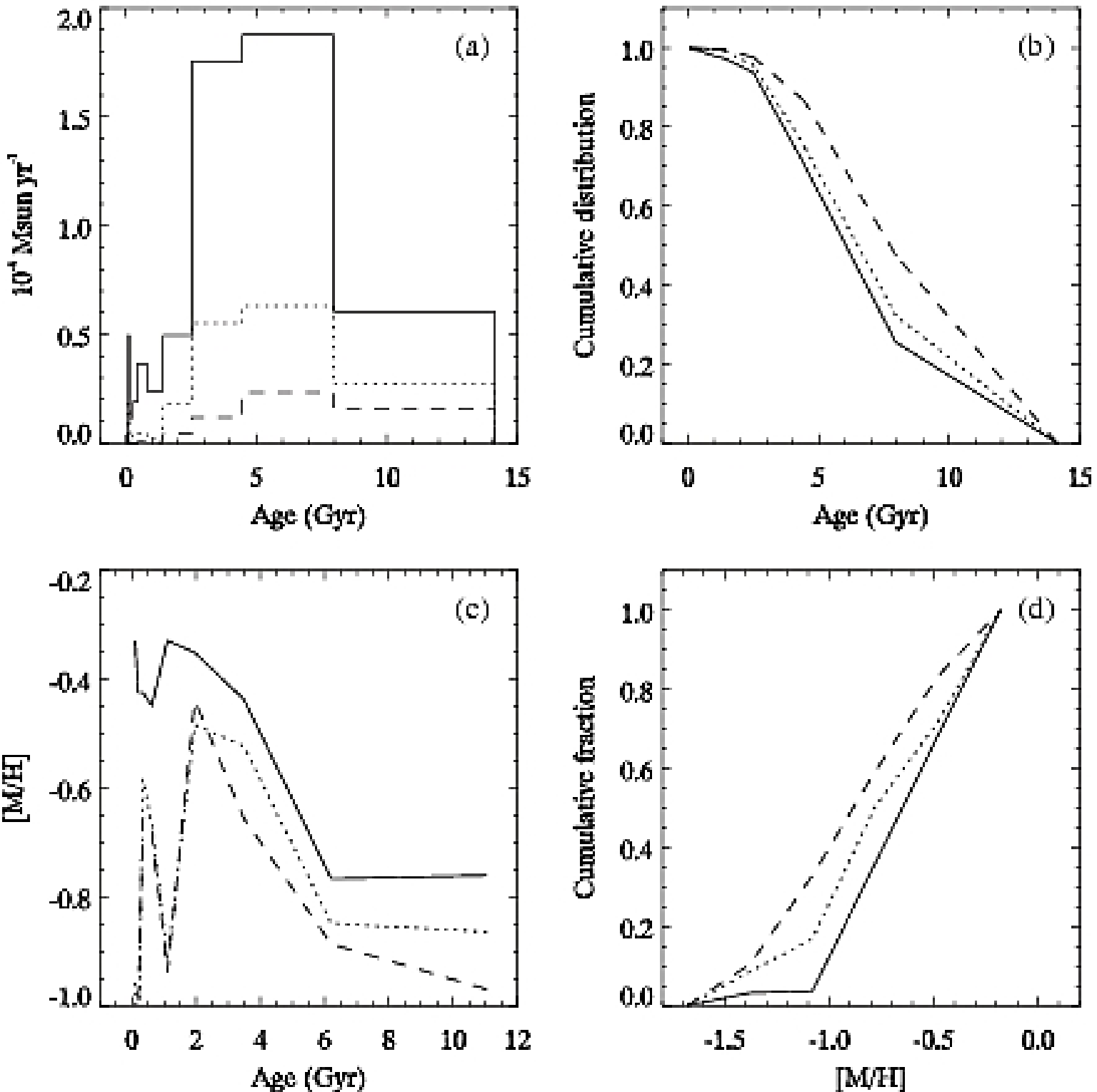}
\caption{Same as Fig.\ \ref{fig:padova} but for the Teramo tracks.}
\label{fig:teramo}
\end{figure*}

\begin{figure*}
\epsscale{1.0}
\plotone{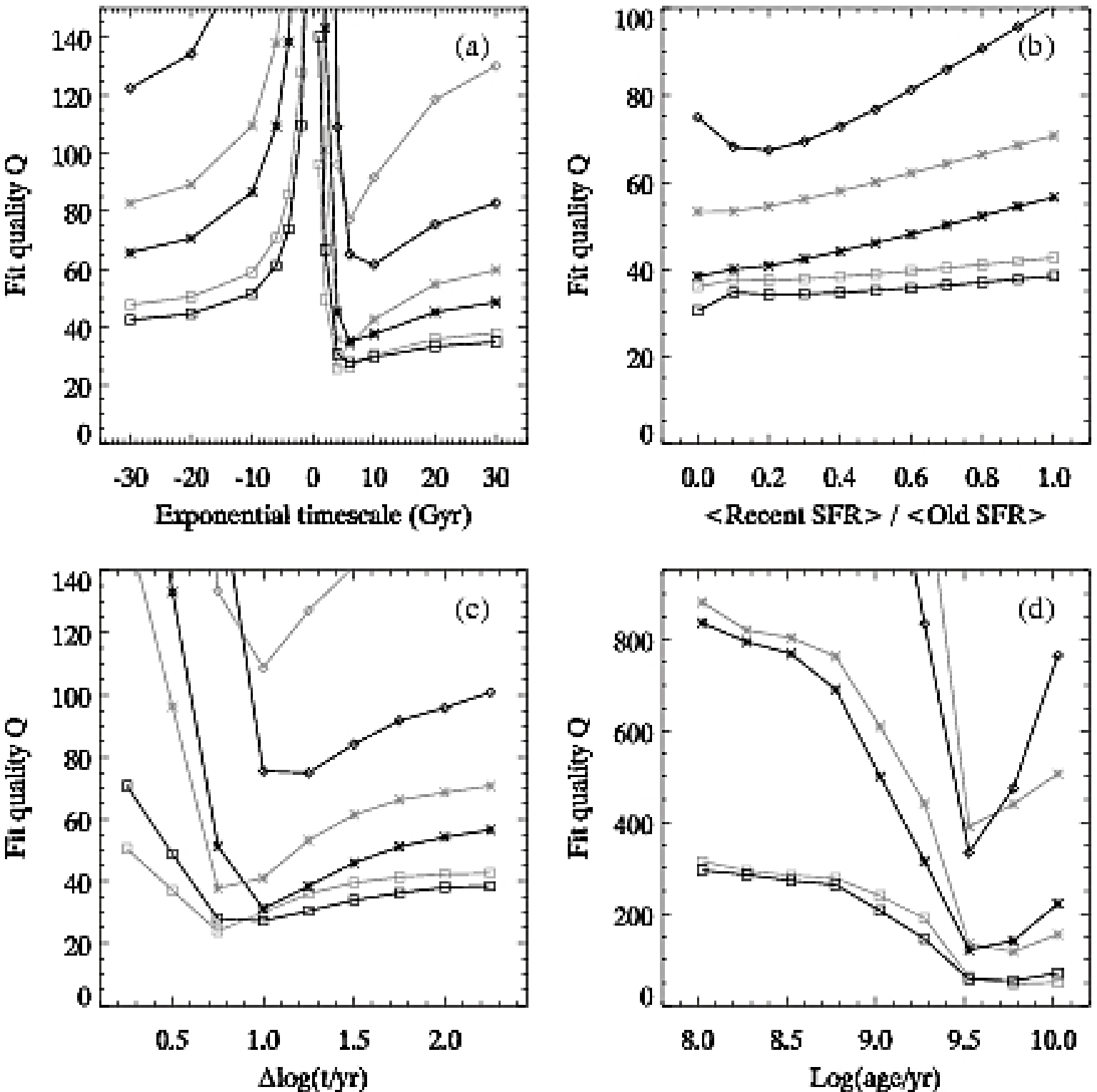}
\caption{Same as Fig.\ \ref{fig:padovaQ} but for the Teramo tracks.}
\label{fig:teramoQ}
\end{figure*}

The new solutions are displayed in 
Figures \ $\ref{fig:A1_padova} - \ref{fig:A3_padova}$.  
The largest
discrepancies occur in the red clump (RC) 
and HB where the model overpredicts
the number of stars.  Disagreements in the RC and HB 
are common (e.g., Dolphin et al.\ 2003; Wyder 2001, 2003)
due to uncertainties in the stellar evolutionary tracks.
In A1 and A2, the model also underpredicts the
number of stars just above the RC possibly
indicating a problem with the AGB bump.
In A3 these discrepancies are not
as significant which could reflect a less complex SFH
relative to A1 and A2.  

Table 1 gives the fit quality and mean distance and 
extinction for each field together with their $1\sigma$
uncertainties.
Table 2 lists the mean age, metallicity, and V-band
mass-to-light ratio ($M/L_V$) with their $1\sigma$
uncertainties.  Tables 3 $-$ 6 provide the SFH, 
age-CDF, AMR, and Z-CDF of the solutions.

To facilitate comparison between the three fields we
show their results together in Figure \ref{fig:padova}.
In each graph, A1 is the solid line, A2 the dotted line,
and A3 the dashed line.
The error bars have been omitted for clarity but they are
the same as in Figs.\ $\ref{fig:A1_padova} - \ref{fig:A3_padova}$.  
The SFHs are qualitatively
similar which is not surprising since the CMDs are similar, too.
Field A1 shows an enhancement in the SFR during the period
2.5 $-$ 4.5 Gyr ago by a factor of $\sim 3 - 4$ over the mean SFR
at older ages.  Because of the large age bins employed, this
does not necessarily mean the true SFH peaked at these
exact ages (see Appendix).
This is followed by a decline in the SFR toward younger ages 
until 250 Myr ago at which time the SFR increases.  This
might suggest a burst of star formation (SF) in the last 250 Myr but the
youngest few age bins are dominated by small number statistics
so the SFRs are not well constrained (see Appendix).
More importantly, the SFR in the youngest few age bins could be 
overestimated if there are stars present in the data with 
ages $\lesssim 80$ Myr, the youngest age covered in the 
synthetic CMDs.  For example, if the true SFR over the past
$\sim 140$ Myr has been constant then the SFR in the youngest bin
could be overestimated by a factor of $\sim 2$ to account for
the stars formed over the last 80 Myr.  Because of the
correlations between age bins some of this overestimation
may leak into nearby bins.  Hence, the SFRs in the youngest $1 - 3$
bins should be viewed as upper limits.

A2 shows a similar behavior but the mean SFR at ages $< 5$ Gyr
is smaller relative to older ages.  The enhancement at $\sim 3$ Gyr
in A1 and A2 is almost nonexistent in A3.  Indeed, the SFH
of A3 is consistent with a constant SFR at ages $> 2.5$ Gyr.
The age-CDFs 
demonstrate more clearly that the fraction of stars 
formed prior to 4.5 Gyr ago increases from $\sim 50\%$ in 
A1 to $\sim 75\%$ in A3.  The mean age of all stars ever 
formed increases from $6.09^{+0.59}_{-0.67}$ to 
$7.99^{+0.86}_{-0.98}$ Gyr.
The mean metallicity decreases from $-0.77^{+0.11}_{-0.12}$
to $-0.93^{+0.19}_{-0.16}$.

In all three fields, the mean 
metallicity of forming stars increases in the oldest three
age bins and then fluctuates wildly at younger ages.
The fluctuations are smallest in A1 and largest in A3 
suggesting they are the result of small number statistics
at ages $\lesssim 2$ Gyr which contribute only $\lesssim 5\%$ 
to all stars in A2 and A3 (see Appendix).  
It is more instructive to
average the metallicity at these young ages.  When we
do this we find that the mean metallicity for ages 
$\leq 2.5$ Gyr is $-0.53$, $-0.67$, and $-0.69$ with a standard
error in the mean of $\approx 0.07$ dex.  Hence, 
we can say with $95\%$ confidence that the mean
metallicity of A1 at young ages is higher than that in
A2 and A3 but A2 is consistent with A3.

As a consistency check on the solutions we can explore the parameter
space by hand using the synthetic CMDs with
$(m-M)_0 = 24.60$ and $A_V = 0.20$.  Panel (a) of
Figure \ref{fig:padovaQ} shows what happens to the 
fit quality, $Q$, when we adopt
an exponentially decreasing or increasing SFH and
vary the timescale, $\tau$.  Positive (negative) 
timescales correspond to a
decreasing (increasing) SFR since formation time.  The
diamonds, asterisks, and squares represent A1, A2, 
and A3, respectively.  For each field we have set 
$\tau = 1, 2, 4, 6, 10, 20$, and 30 and we normalize the
model CMD to have the same total number of stars as the data
CMD in the fitted region.  
The black lines
correspond to synthetic CMDs with metallicities
$\rm [M/H] = -0.8$ to $-0.5$ while the gray
lines show the effect of using the next lowest
metallicity bin, $\rm [M/H] = -1.3$ to $-1.0$.

The lowest $Q$ values in panel (a) are much larger
than the global best-fit SFHs suggesting that the true SFH 
(averaged over our age bins) has not exactly followed
an exponential throughout M33's lifetime.
Nevertheless, the preferred timescale decreases from $\sim 20$ Gyr
in A1 to $\sim 6$ Gyr in A3.  
The difference in fit quality between exponentially increasing
and decreasing SFHs with the same timescale is mainly 
sensitive to the mean SFR over the past $\sim 1$ Gyr.  Ages 
$\lesssim 1$ Gyr contribute mostly to the CMD at colors
$(V-I) \lesssim 0.5$ while the opposite is true
for older ages.  Exponentially increasing SFHs have many
more stars at these blue colors than are observed 
in the data CMDs.
The fact that exponentially
decreasing SFHs are preferred indicates
that the average SFR in the past $\sim 1$ Gyr has been lower than 
at older ages.  

We explore this fact further in panel (b) which shows
the ratio between the average recent SFR (ages $< 794$ Myr)
to the average SFR at older ages.  The SFR in both
regimes has been set constant and the ratio was varied in
increments of 0.1.  In A1 there is an unambiguous minimum at 
$0.2 - 0.3$, in A2 the minimum is less significant
but occurs at $0.1 - 0.2$, and in A3 there is little constraint 
although the minimum occurs at 0.0.  This supports the idea
that, at least in A1 and A2, the mean SFR over the 
past $\sim 1$ Gyr has been lower than at older ages.  
The precise value of the ratio could depend on the
shape of the true IMF relative to that of the model.  
For example, a model IMF with a single slope that is 
steeper (shallower) than in the data 
will cause an overestimate (underestimate) 
of the recent SFR because more young, 
high-mass stars will be needed to match the observations
(Dolphin 1997).  

Panel (c) explores
the duration of star formation for a constant SFR starting
at $\rm log(t/yr) = 10.15$ (14.1 Gyr).  The duration is
successively increased in steps of 0.25 dex.  We see that
the optimal duration is 1.5 dex in A1 and 1.25 dex in A2 and A3.
This further supports the case for a longer era of star 
formation in A1 than in A3.

In a similar manner we can also see what particular age
range is preferred by comparing each individual synthetic CMD
to the data CMD.  We plot
the results of this exercise in panel (d).  
This plot demonstrates that the synthetic CMD centered on 
log(age/yr) = 9.525 (age range
$\approx 2.5 - 4.5$ Gyr) generally provides the closest match to the
observed CMDs with a modest dependence on metallicity due 
to the age-metallicity degeneracy.
Interpolating smoothly between the points by eye, it appears that
the preferred age increases from A1 to A3 providing yet 
another confirmation that the mean age increases throughout
the fields.
Since the lowest $Q$ values
are still very high we can conclude that SF has occurred over
timescales longer than those covered in each of the 
synthetic CMDs alone.  Overall, 
Fig.\ \ref{fig:padovaQ} makes it easier to
understand the global best-fit solutions for A1 and A2 
which show an enhancement in the SFR at $2.5 - 4.5$ Gyr 
with approximately constant SFR at other ages and a 
drop in the average SFR over the past 1 Gyr.

\subsection{Teramo tracks}

Figures $\ref{fig:A1_teramo} - \ref{fig:A3_teramo}$ display 
the global solutions obtained
with the Teramo tracks.  There are several
striking differences between the Teramo and Padova solutions.  
First, the enhancement at intermediate ages occurs over
two contiguous age bins from $\sim 2.5 - 8$ Gyr
rather than just one bin.
In A1, the strength of the enhancement is about
$70\%$ smaller since it lasts for a longer timespan
and the total number of stars 
must be conserved.  In addition, a larger fraction
of stars formed by 5 Gyr ago -- $\sim 70\%$ as opposed
to $\sim 50\%$ for the Padova tracks.

Another difference between the Teramo and Padova solutions
is that the RC area in the Teramo model
CMD provides a better fit to the data.  Interestingly, 
though, the model has an extended blue HB which is
not as obvious to see in the data.  
The Teramo models have a noticably more extended 
HB at old ages than the Padova models 
(Gallart et al.\ 2005).
Although the extended HB does not appear to significantly
constribute to the residuals, 
the large width of the oldest age bin may have 
forced the model to include very old stars 
($\gtrsim 12$ Gyr) that might not be in the data.  
We re-ran the fitting routine but changed
the oldest age bin to have a width of
0.15 dex so the oldest stars included
in the model were 11.2 Gyr old.  As expected, the
HB morphology of the resulting fits was redder and
similar to the Padova results in the previous
section but the fit qualities were no better
and the SFH was not significantly changed.

One notable difference between the Teramo
model CMDs and data CMDs is that the models contain an
excess of stars on the RGB.  
This is a known issue with the Teramo models which
predict RGB lifetimes that are too long.
This results in an RGB luminosity function with too
many stars when compared to other sets of models and
GGCs (Gallart et al.\ 2005).
Since this discrepancy
does not affect the color of the RGB it probably
has little impact on our results.

We note that the best-fit distance modulus
is $\sim 0.1$ mag greater for the Teramo tracks
than for the Padova tracks yet 
the best-fit extinction values are similar.  This underscores
the fact that distances and extinctions 
obtained with this technique are
subject to systematic errors in 
the stellar evolutionary tracks.  These errors can
depend on filter, age, and metallicity among other
factors (Wyder 2003).

The Teramo solutions are overplotted on each other
in Figure \ref{fig:teramo}.  
Despite the differences from the Padova solutions,
they exhibit similar trends between
the three fields.  
The strength of the enhancement at
intermediate ages decreases relative to the SFR
at older ages.  Consequently, the mean age
of the fields increases from 
$6.50^{+0.46}_{-0.51}$ Gyr in A1 to 
$8.09^{+0.97}_{-1.24}$ Gyr in A3.  The mean metallicity decreases
from $-0.66^{+0.11}_{-0.11}$ to $-0.89^{+0.18}_{-0.18}$.  
The apparent lack of evolution in the AMR
between the two oldest bins does not necessarily
weaken the validity of the model nor does it
necessarily mean there was no change in the
true AMR.  As shown in the Appendix, variations between
adjacent bins must be considered with caution.

Figure \ref{fig:teramoQ} shows again that the Teramo
tracks predict similar trends between the fields 
as the Padova tracks although
some specific details are different.  The preferred exponential
timescale decreases from 10 Gyr in A1 to $4-6$ Gyr in A3.  
The ratio of recent SFR to past SFR is $0.1-0.2$ in A1
and consistent with zero in A2 and A3.  The duration of
star formation decreases from $1-1.3$ dex in A1 to $0.7-1.0$ dex
in A3.  Finally, the preferred age increases from log(age/yr) $= 9.5$
in A1 to log(age/yr) $= 9.8$ in A3.

\section{Discussion}
\label{sec:disc}

There have been few studies of the resolved stellar 
populations in the outer regions of late-type spiral galaxies.
Davidge (2003) imaged the outskirts of M33 and NGC 2403 with the
Gemini Multi-Object Spectrograph (GMOS) on Gemini North 
covering deprojected radii of $14 - 17$ kpc and 
$7 - 19$ kpc, respectively.
He found evidence for bright RGB and AGB
stars well mixed throughout the observed fields and interpreted
them as evidence for in-situ SF at
intermediate ages.  Bland-Hawthorn et al.\ (2005) 
used the GMOS on Gemini South
to observe the outskirts of another late-type spiral, NGC 300.
Using star counts in the $r'-$band, they found the 
exponential disk to extend out to $R_{dp} = 14.4$ kpc 
or 10 optical scale lengths.  Seth et al.\ (2005)
studied the vertical distribution of resolved stars in six low-mass
spiral galaxies and found the stellar component to extend
up to 15 scale heights.  
Our results complement these
other studies well and together they suggest that the 
stellar disk populations of late-type spirals commonly extend out 
to large distances.  However, just because the stellar
surface density or light distribution extends to these
large distances does not necessarily mean the stars belong
to a kinematic disk population (i.e., rotationally supported
with ordered motion).

\begin{figure}
\epsscale{1.0}
\plotone{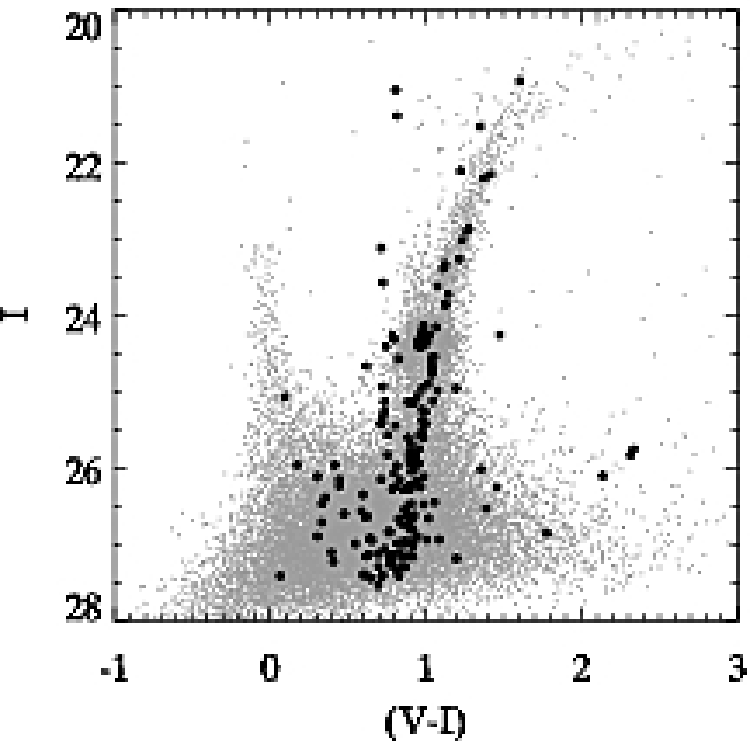}
\caption{CMD of field A1 (gray points) with CMD of
Terzan 7 overplotted (black points).}
\label{fig:ter7}
\end{figure}

An empirical check on our results is displayed in 
Figure \ref{fig:ter7}.  This figure shows the CMD of field
A1 as gray points with the CMD of globular cluster Terzan 7
as black points.  The Ter 7 data come from 
Sarajedini \& Layden (1997) and we have plotted only stars 
within the central $80\arcsec$ of the cluster center.
Most of the black points bluer than $(V-I) \sim 0.8$ are
Galactic field stars.  Ter 7 belongs to the Sagittarius
dwarf galaxy and is estimated to have a metallicity
[Fe/H] $= -0.82 \pm 0.15$ and to be $\sim 6$ Gyr younger
than 47 Tuc (Sarajedini \& Layden 1997).  The RC and RGB of Ter 7
closely match those of M33 confirming our general result
that M33's outskirts are $6-8$ Gyr old with metallicities
of $\sim -0.7$ to $-0.9$.

\begin{figure}
\epsscale{1.0}
\plotone{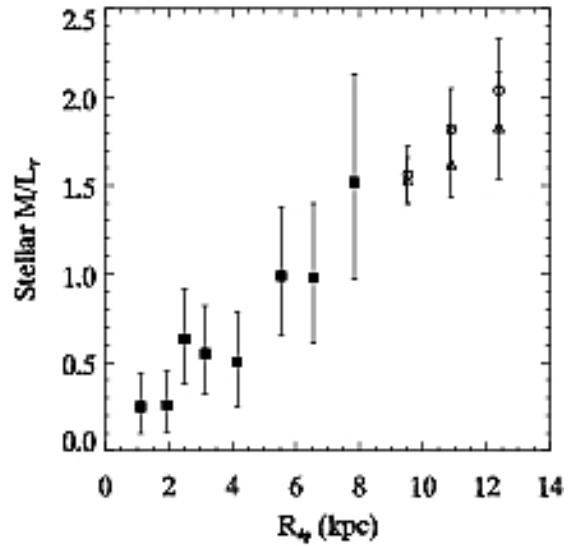}
\caption{V-band stellar mass-to-light ratio in M33.  Squares
represent the data based on PNe kinematics (Ciardullo et al.\ 2004).
Circles and triangles correspond to the SFH results for fields
A1 $-$ A3 using the Padova and Teramo models, respectively.}
\label{fig:MLratio}
\end{figure}

Ciardullo et al.\ (2004) conducted a photometric and
spectroscopic survey of planetary nebulae (PNe) throughout
M33.  They estimated the disk mass surface density
under the epicyclic and isothermal disk approximations 
by combining the vertical velocity dispersions of the
PNe with published optical surface photometry and gas mass 
surface densities.  To account for extinction they adopted the
simple exponential model of Regan \& Vogel (1994) 
with a central $A_V$ of 0.9.
Ciardullo et al.\ found that $M/L_V$ 
of the stellar component increases
from $\sim 0.3$ to $\sim 1.5$ over the face of the disk.
We show their results as squares in Figure \ref{fig:MLratio}
after transforming to a Galaxy-M33 distance 
of 867 kpc  (Paper I; Galleti et al.\ 2004).
The $M/L_V$ values for fields A1 $-$ A3 derived from
our SFH analysis are shown as circles (Padova) and
triangles (Teramo).  

This provides a nice consistency check on our SFH results
because $M/L_V$ depends heavily on age and to a lesser
extent on metallicity.  
For instance, the Padova synthetic CMDs covering metallicities
from $-0.8$ to $-0.5$ have an
$M/L_V$ that increases with age from 0.16 to 3.57.  
The relative proportions of
different ages in the SFHs affect both the
normalization of $M/L_V$ and its change with radius.
The agreement between the two sets of data thus
provides independent support that our SFH results
contain a reasonable mix of ages and metallicities.

More fundamentally, the
agreement supports the IMF we used in calculating
the SFHs.  Recall that the low-mass exponent ($M \leq 0.5\ M_{\sun}$) 
was $-1.35$.  If we steepen it to $-2.00$ then $M/L_V$
increases by $\sim 20\%$ without affecting the SFHs.
A flatter low-mass slope of $-0.70$ would decrease
$M/L_V$ by $\sim 10\%$.
Changing the IMF slope at higher masses could affect the
SFHs and resulting $M/L_V$ in a non-trivial way.
However, it is unlikely that we just happened to
pick a particular form of the IMF which yields a similar 
$M/L_V-R_{dp}$ relation as the PNe kinematics.
Therefore, it seems that the IMF in M33's outskirts
is similar to the Galaxy's or is at least shallower
than a Salpeter form at the lowest masses.

Taken at face value, Fig.\ \ref{fig:MLratio} also suggests
that the mean age of M33's {\it entire} stellar disk 
increases with radius.  All else being equal,
the correlation between age and $M/L$ is 
one-to-one for simple stellar populations but late-type galaxies
like M33 are characterized by SF at a wide variety of ages.  In systems
where SF has occurred in the last $\sim 1$ Gyr, the light
from the most massive, youngest stars can completely overwhelm the
light from older stellar generations thus weakening or
even reversing the correlation between $M/L$ and mean age.
This is because
while the youngest stars contribute the majority
of the light they have only a small effect on the
age averaged over a Hubble time.  Therefore, it may be
premature to extrapolate the positive age gradient in M33's outer
disk to its inner disk.
In future work, we will examine the SFH of M33's inner disk to 
shed further light on this issue.

An increasing mean age in M33's disk is apparently at
odds with the inside-out scenario of galaxy formation.  There is a wide
body of theoretical and observational evidence to support 
the inside-out scenario.
It is a standard prediction of hierarchical disk galaxy
formation models with cold dark matter 
(e.g., Fall \& Efstathiou 1980; Mo et al.\ 1998).
The sizes of disk galaxies are observed to decrease
with redshift in rough accordance with 
model predictions (Ferguson et al.\ 2004).
Furthermore, galactic disks in the local Universe generally become
bluer with increasing radius which is usually interpreted
as a decreasing mean age (e.g., de Jong 1996; Bell \& de Jong 2000).  

Does the inside-out build-up of dark matter halos
necessarily result in {\it stellar disks whose
mean ages at the present epoch} decrease with radius?
The question is difficult to answer because it requires
incorporating the highly uncertain physics of gas cooling, 
star formation, feedback, and the effects of an ionizing
background into the results of
hierarchical cosmological simulations 
(Silk 2003).  Most theoretical predictions for
the run of mean age with radius come from simplified 
analytic or semi-analytic treatments
of the baryonic physics 
(e.g., Moll\'{a} \& D\'{i}az 2005; Naab \& Ostriker 2006).  
These studies do predict the mean stellar age to decrease with radius
under the inside-out hierarchical framework.
However, the treatments of the processes may 
break down in the outskirts of disks or
the various assumptions and 
neglected processes in these predictions could 
change the results.
Perhaps the initial build-up of the disk proceeds quickly
so that after a few Gyr there is a negative age gradient
but subsequent processes like gas infall, outflow, 
and viscous radial flows reverse this gradient.  Alternatively,
star formation could begin in an inside-out fashion but
it could be truncated outside-in producing a positive
age gradient because the inner regions would be forming
stars for a longer period of time.

On the observational side, 
the nature, ubiquity, and interpretation of negative
disk color gradients is not entirely clear.
MacArthur et al.\ (2004) examined optical and
near-IR color gradients for a large sample of galaxies
and derived luminosity-weighted 
mean age and metallicity profiles from stellar
population synthesis models.
They found evidence for a radial 
dependence of age gradients in the sense that the
inner regions showed generally steeper gradients
than the outer regions.  In addition, some galaxies displayed
inflection points in their age gradients.
Taylor et al.\ (2005) found a morphological dependence of
color gradients such that early-type systems tended to get
bluer with radius whereas
late-type spiral, irregular, peculiar, and
merging galaxies tended to get redder with increasing radius.
They attributed this to mergers, accretions, and interactions
triggering radial inflows of gas and centrally concentrated
starbursts.  They also found that galaxies with faint
absolute B-band magnitudes
were somewhat more likely to get redder with radius than their
brighter counterparts.  
On the other hand, Jansen et al.\ (2000) found 
no color gradient dependence
on morphological type but an even stronger trend with B-band magnitude.
They concluded that star
formation tends to occur in the outer regions of luminous
galaxies but in the inner regions of fainter systems.

How does M33 compare with the results of the aformentioned studies?
Because of its large angular extent in the sky and low
surface brightness, M33's color gradients are not well known.
Guidoni et al.\ (1981) carried out $UBVRI$ photoelectric measurements
of the central $13\arcmin$ and found the disk colors to change
little with radius except for $(U-B)$ which is heavily
influenced by dust.
The 2MASS Large Galaxy Atlas (Jarrett et al.\ 2003) 
reports $(J-K_S)$ as increasing
from $\approx 0.8$ to 1.0 over the inner $9\arcmin$.
Regan \& Vogel (1994), on the other hand, 
found $(J-K)$ to {\it decrease}
from $\approx 1.0$ to 0.8 over the same region.  
The cause of the discrepancy
is not clear but it could arise from uncertainties 
in the sky subtraction.  In any case, 
our SFH results predict instegrated colors of 
$(V-I) \approx 1.0$, $(B-I) \approx 1.7$, 
and $(J-K) \approx 0.8$ in M33's outer disk.  
These values are
in reasonable agreement with the published measurements but it
would be worthwhile to update and extend the radial coverage of
M33's surface photometry.  Such data could provide independent
constraints on SFH analyses similar to our own.

\begin{figure*}
\epsscale{1.0}
\plotone{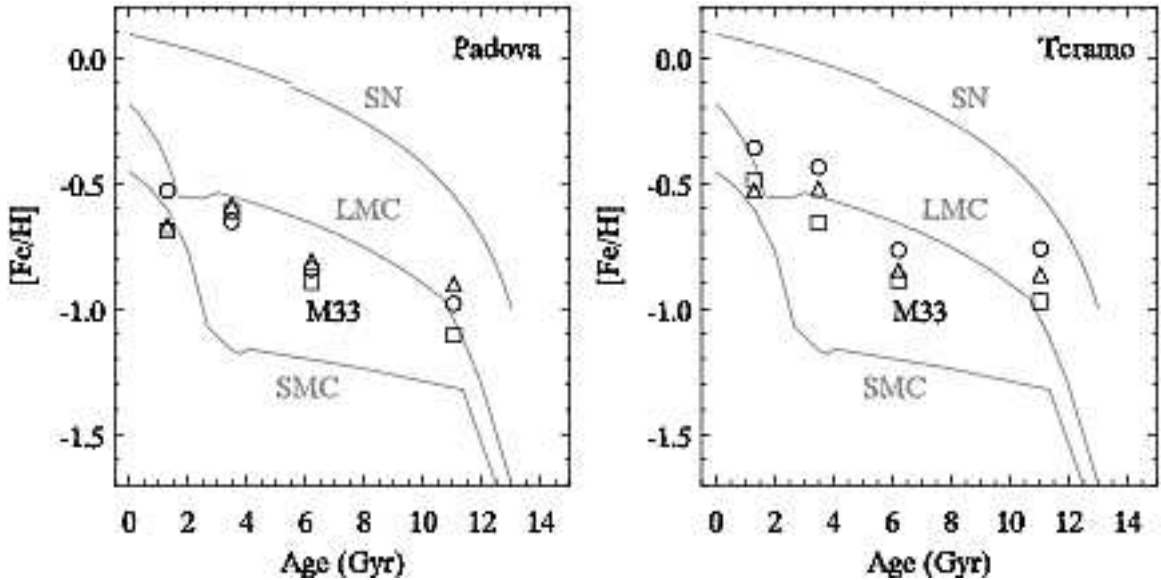}
\caption{AMR of M33 compared to the SN, LMC, and SMC shown
as gray lines.  Fields A1, A2, and A3 correspond to circles, triangles,
and squares, respectively.  The left panel shows the Padova results
while the right panel shows the Teramo results (see text for details).}
\label{fig:amr_lmc}
\end{figure*}

It is interesting to compare our results for M33's AMR to that
of other well-studied systems.  In Figure \ref{fig:amr_lmc}, 
the gray solid lines show the
AMRs of the Small and Large Magellanic Clouds (SMC and LMC, or MCs), 
and the Solar neighborhood (SN).
For the MCs we have used the bursting 
models of Pagel \& Tautvai\v{s}ien\.{e} (1998)
which include inflow and non-selective galactic winds.
These authors tuned the parameters of their models to match
the observed abundances of MC clusters and supergiants.
The abrupt change in the enrichment rate at $\sim 3$ Gyr is
due to a burst in star formation possibly caused by an interaction
between the clouds.
The SN model is taken from Twarog (1980) who used the abundances
of a large sample of nearby F dwarfs as constraints.   This model 
incorporated an initial metallicity of $-1.0$ and a constant 
SFR and inflow rate over the disk lifetime.

The existence and nature of an AMR in the SN is a matter of 
some debate.  Edvardsson et al.\ (1993) and 
Feltzing et al.\ (2001) found the
AMR to have a large intrinsic scatter ($\sim 0.2$ dex) 
with the oldest stars being both metal-poor and metal-rich.
Subsequent studies have since challenged their results
citing sample selection effects or biases in the age 
determinations as the cause (Garnett \& Kobulnicky 2000; 
Rocha-Pinto et al.\ 2000; Kotoneva et al.\ 2002; 
Pont \& Eyer 2004; Rocha-Pinto et al.\ 2006).  
In any case, we are concerned with the mean overall level
of enrichment rather than the scatter and in this sense the Twarog
AMR agrees well with most other studies (Rocha-Pinto et al.\ 2000).

The points in Fig. \ref{fig:amr_lmc} show the 
AMRs we derived for field A1 (circles),
A2 (triangles), and A3 (squares).  The Padova results are
shown in the left panel while the Teramo results are shown
in the right panel.  Because the AMR at the youngest ages is
dominated by small number statistics, we have averaged
the 6 youngest age
bins which cover ages $\lesssim 2.5$ Gyr.  

This figure demonstrates that the level of enrichment in M33's outer 
disk has been intermediate between the SMC and LMC but perhaps
somewhat closer to the latter.  Indeed, if the SMC and LMC had continued
to evolve quiescently rather than experience bursts at $\sim 3-4$ Gyr
then their present-day metallicities would have been close to $-0.9$
and $-0.5$, respectively.  In that case M33's outer disk 
would have resembled the
LMC even more.  

Finally, our results imply a present-day global metallicity of 
$\sim -0.5$ in M33's outer disk which is in good agreement with the 
results of Urbaneja et al.\ (2005).  These authors 
conducted a detailed spectral analysis of $\sim 10$ B-type
supergiant stars throughout M33 based on non-LTE
model atmospheres including the effects of stellar winds.
They found [M/H] in their stellar sample to decrease from
about 0.0 near M33's nucleus to about $-0.5$ at $R_{dp} = 33\arcmin$
just interior to field A1.

\section{Conclusions}
\label{sec:conc}

We have conducted a detailed analysis of the SFH of M33's
outer regions by modelling the observed CMDs as linear combinations
of individual synthetic populations with different ages and
metallicities.
To gain a better understanding of the systematic errors we
have conducted the analysis with two different sets of stellar
tracks, Padova and Teramo.
The precise details of the results 
depend on which tracks are used but we can make several
conclusions that are fairly robust despite the differences.

Both sets of tracks predict the mean age to increase 
and the mean metallicity to decrease with radius.
When star formation is restricted to age intervals 0.25 dex
wide and global metallicity intervals 0.3 dex wide, then 
ranges centered on ages $3 - 8$ Gyr and metallicities $-1.0$ to $-0.7$ 
are preferred with A1
matching more closely the younger, metal-rich ends of these ranges and
A3 matching more closely the older, metal-poor ends.
If star formation began at the same time in each field then its timescale
has decreased with radius.

\begin{figure*}
\epsscale{0.95}
\plotone{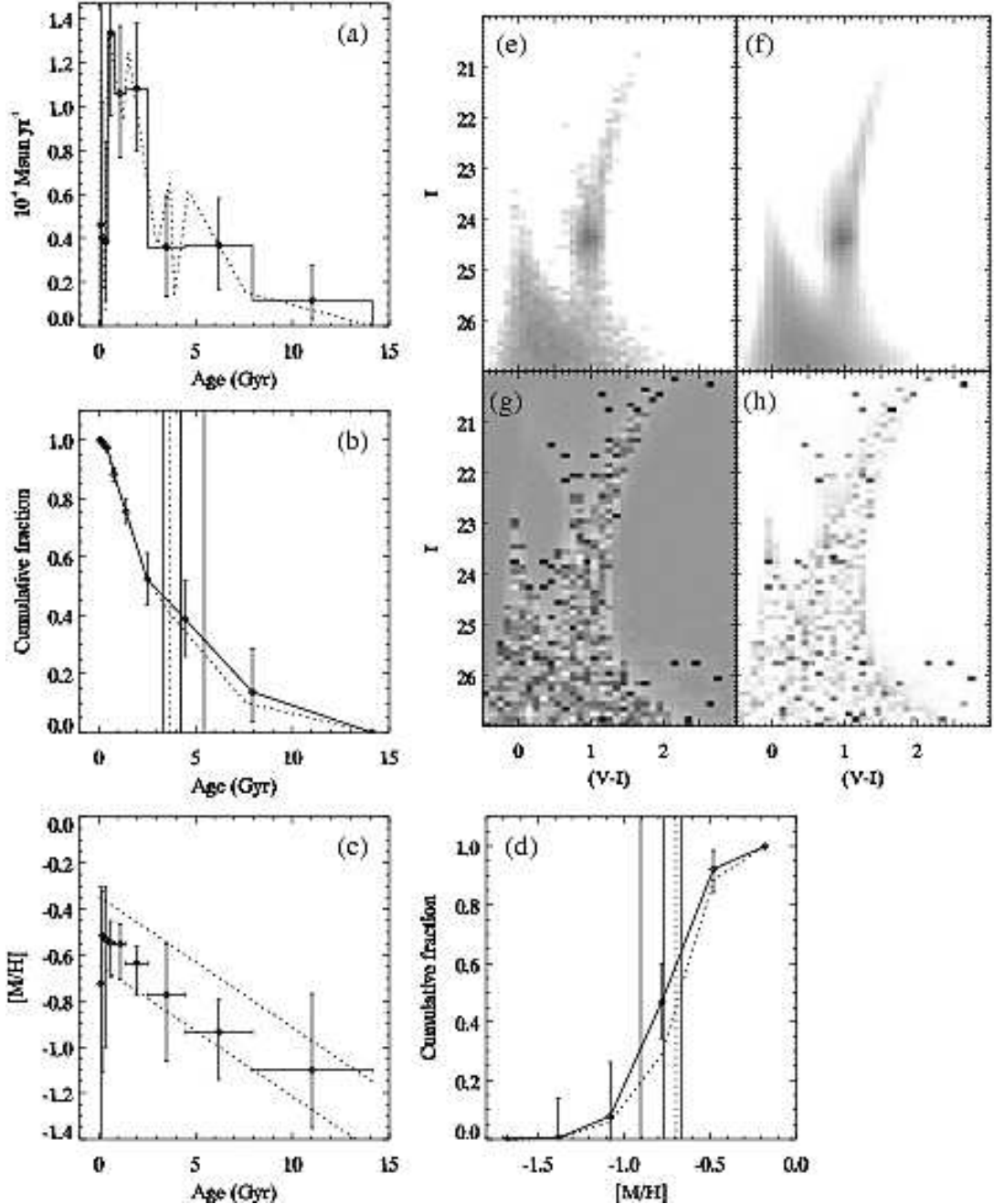}
\caption{Testing the method.  Same panels as 
Fig.\ \ref{fig:A1_padova}.  See Appendix for details.}
\label{fig:test1}
\end{figure*}

\begin{figure*}
\epsscale{0.95}
\plotone{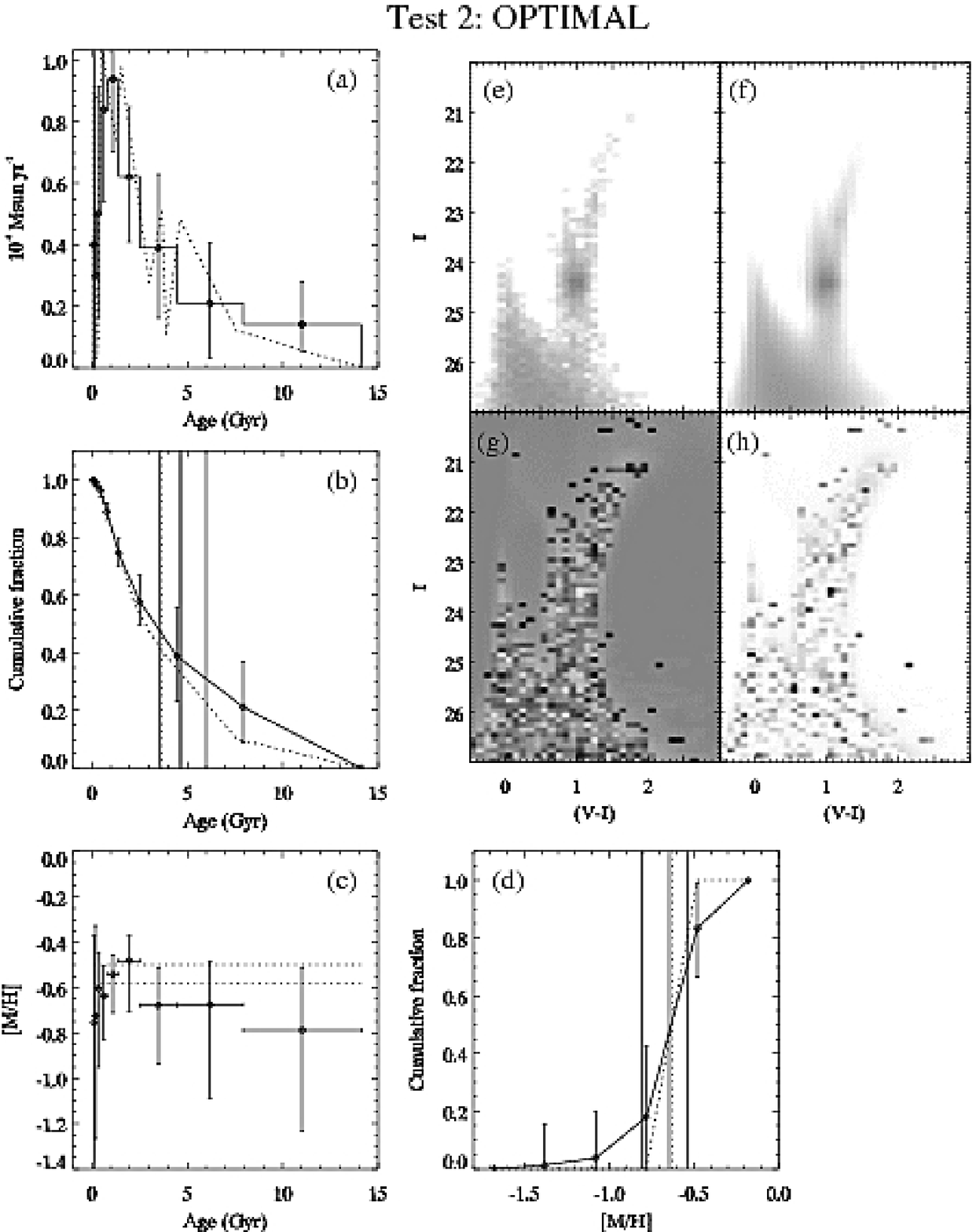}
\caption{Same as Fig.\ \ref{fig:test1}.}
\label{fig:test2}
\end{figure*}

\begin{figure*}
\epsscale{0.95}
\plotone{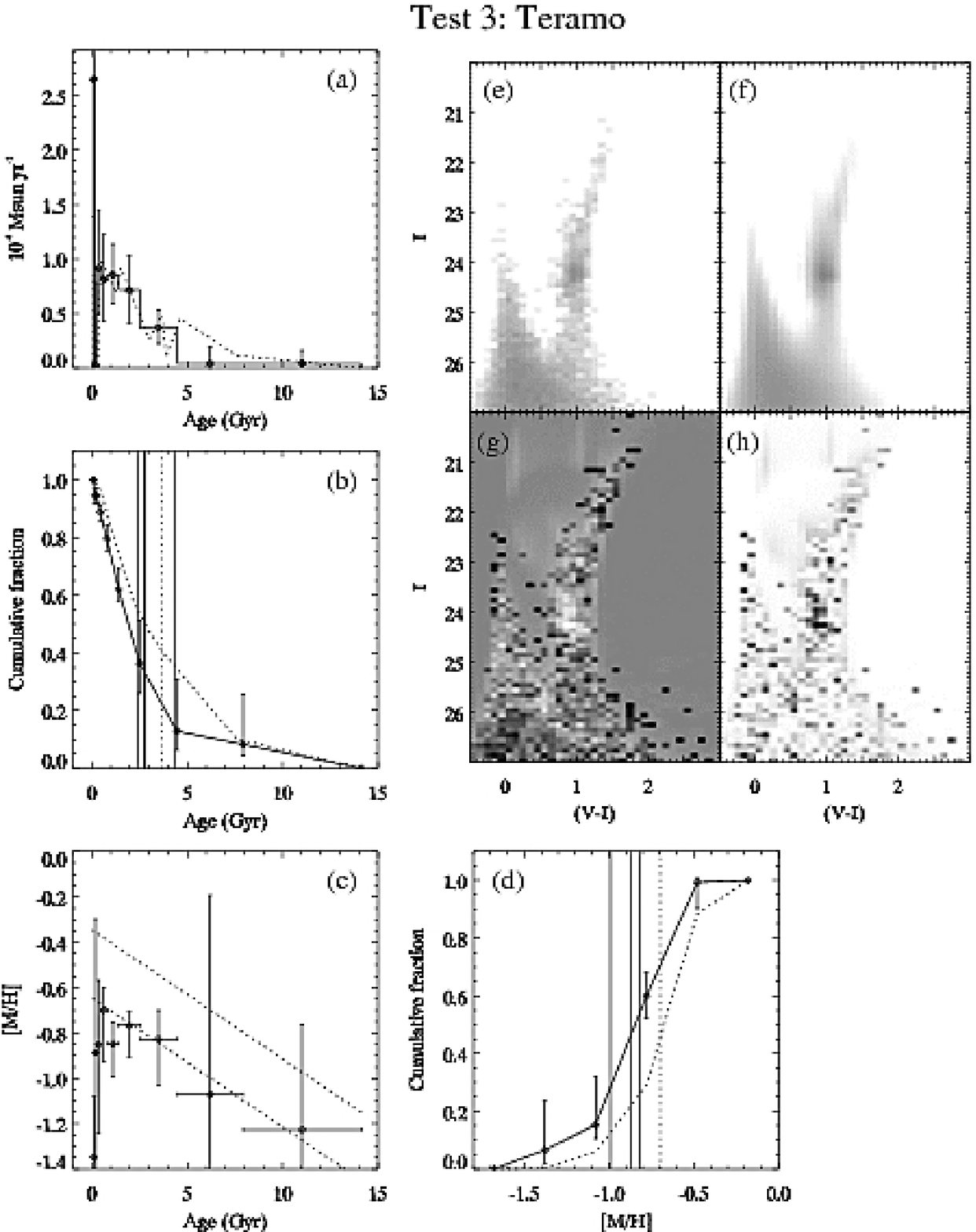}
\caption{Same as Fig.\ \ref{fig:test1}.}
\label{fig:test3}
\end{figure*}

\begin{figure*}
\epsscale{0.95}
\plotone{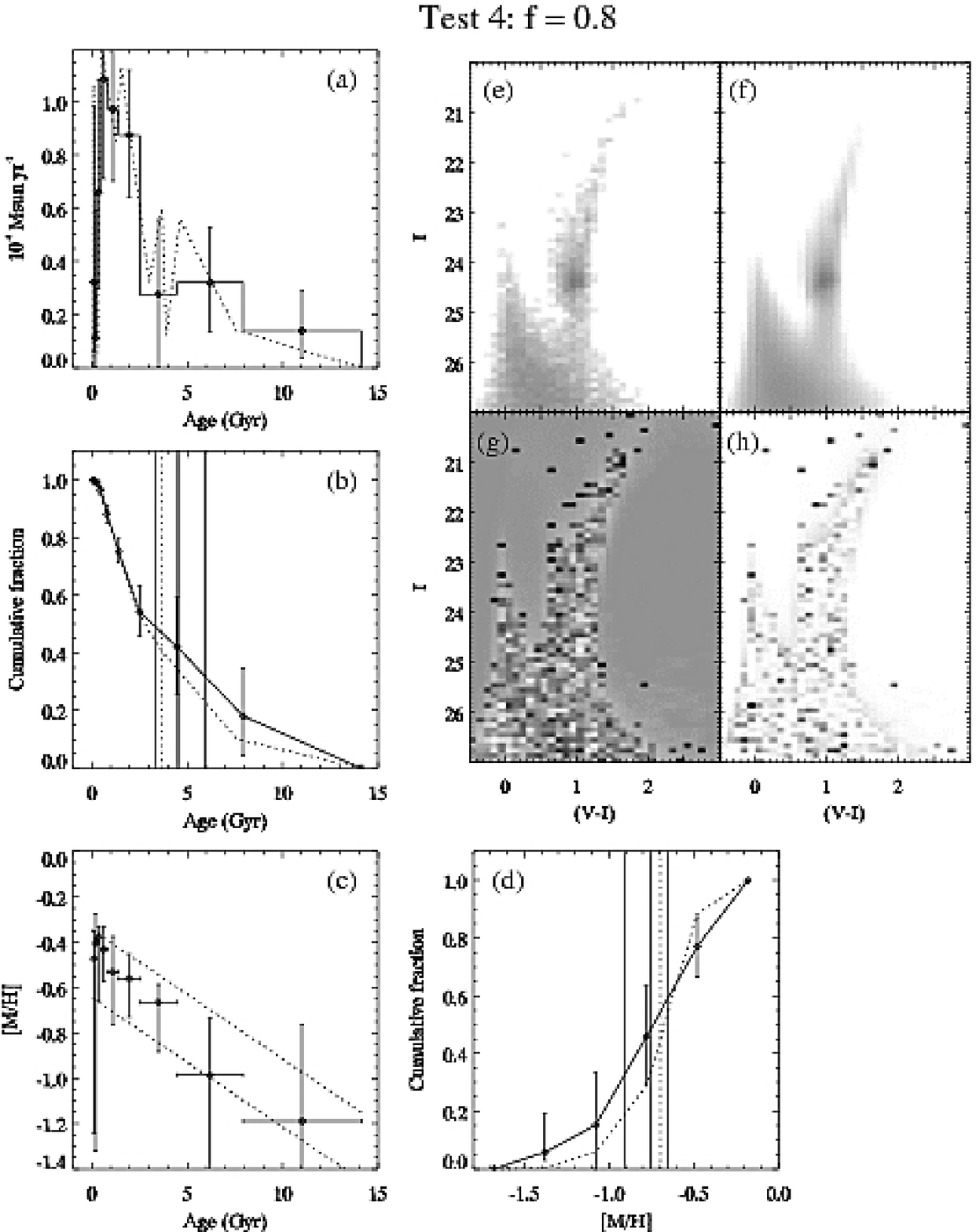}
\caption{Same as Fig.\ \ref{fig:test1}.}
\label{fig:test4}
\end{figure*}

\begin{figure*}
\epsscale{0.95}
\plotone{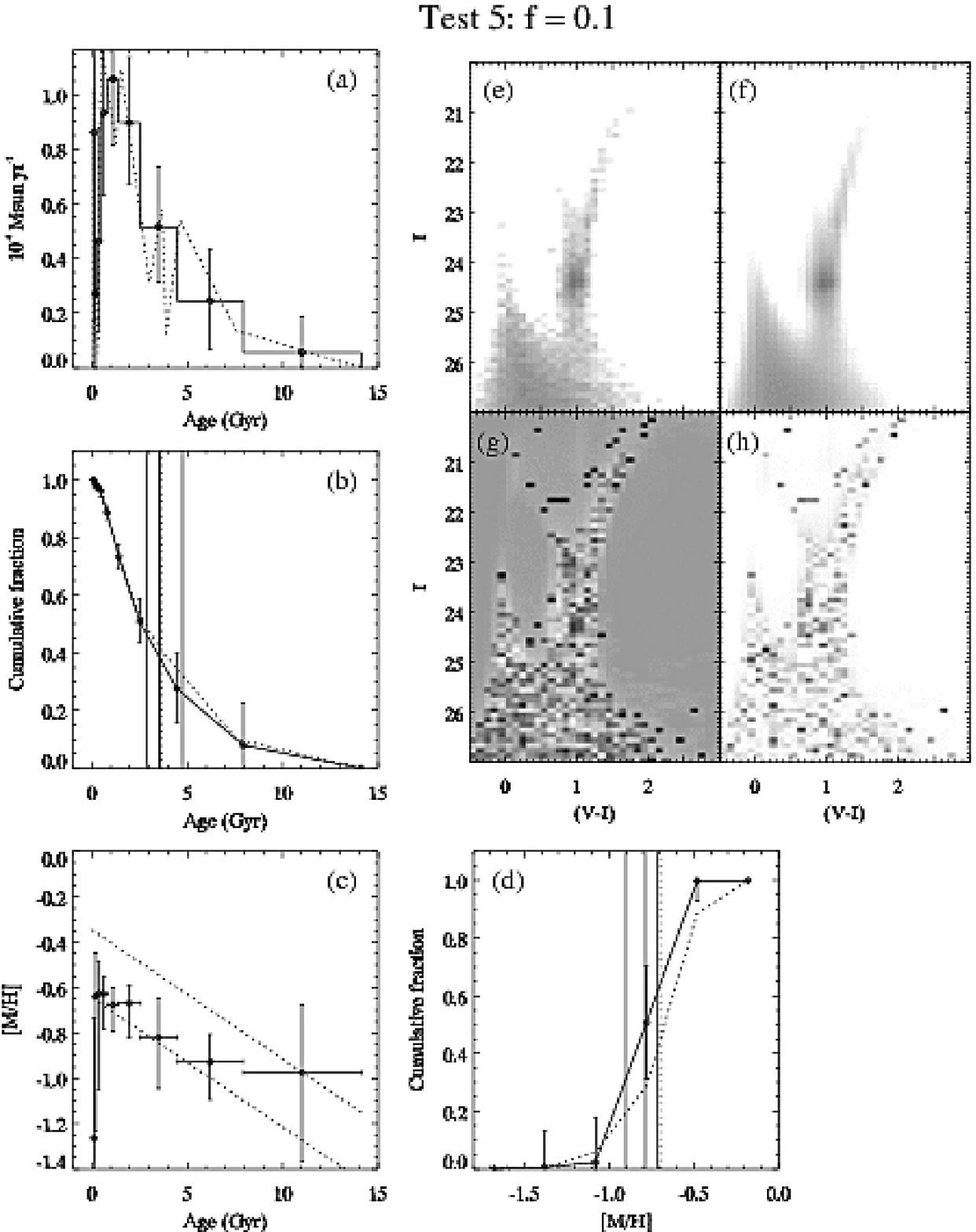}
\caption{Same as Fig.\ \ref{fig:test1}.}
\label{fig:test5}
\end{figure*}

\begin{figure*}
\epsscale{0.95}
\plotone{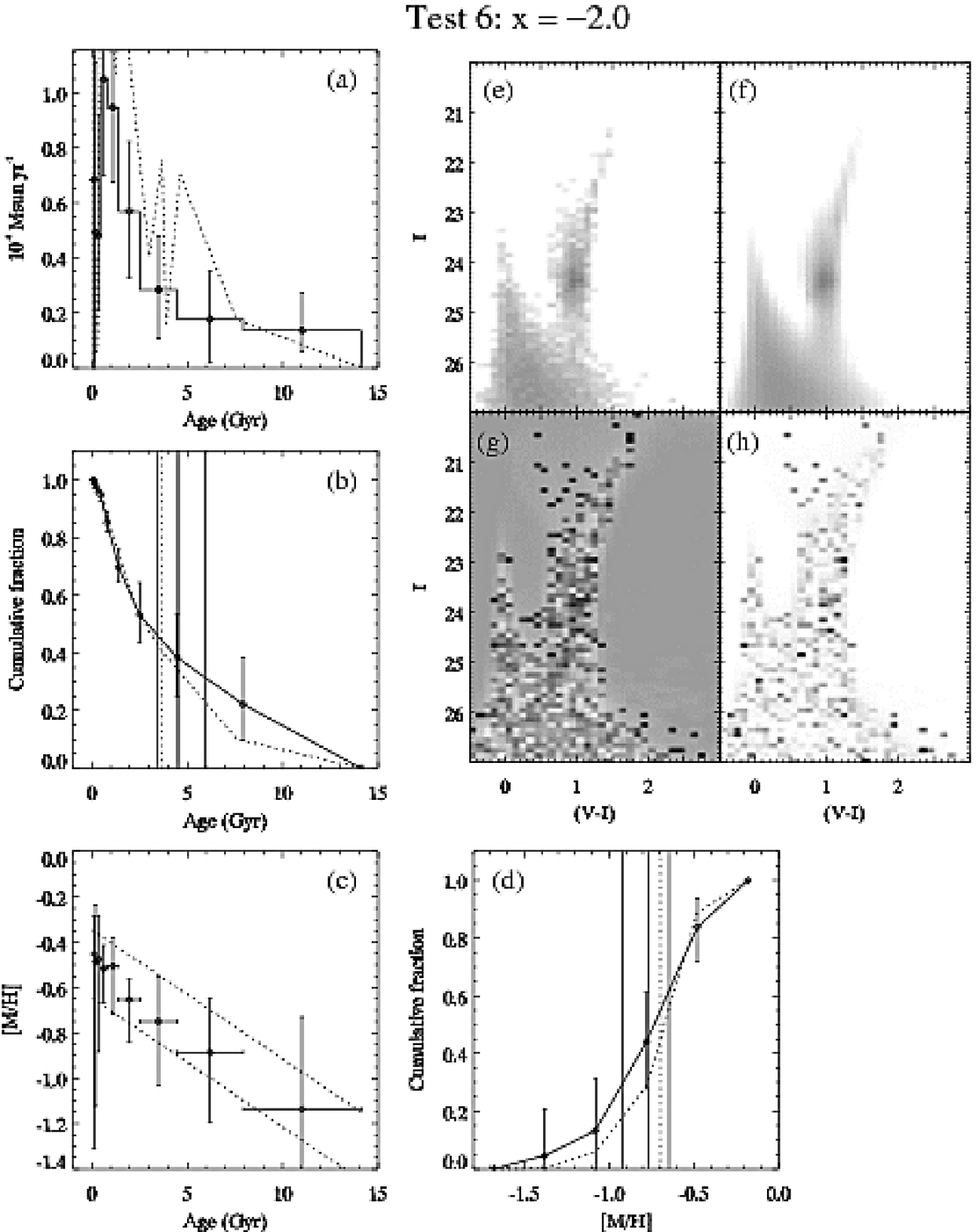}
\caption{Same as Fig.\ \ref{fig:test1}.}
\label{fig:test6}
\end{figure*}

\begin{figure*}
\epsscale{0.95}
\plotone{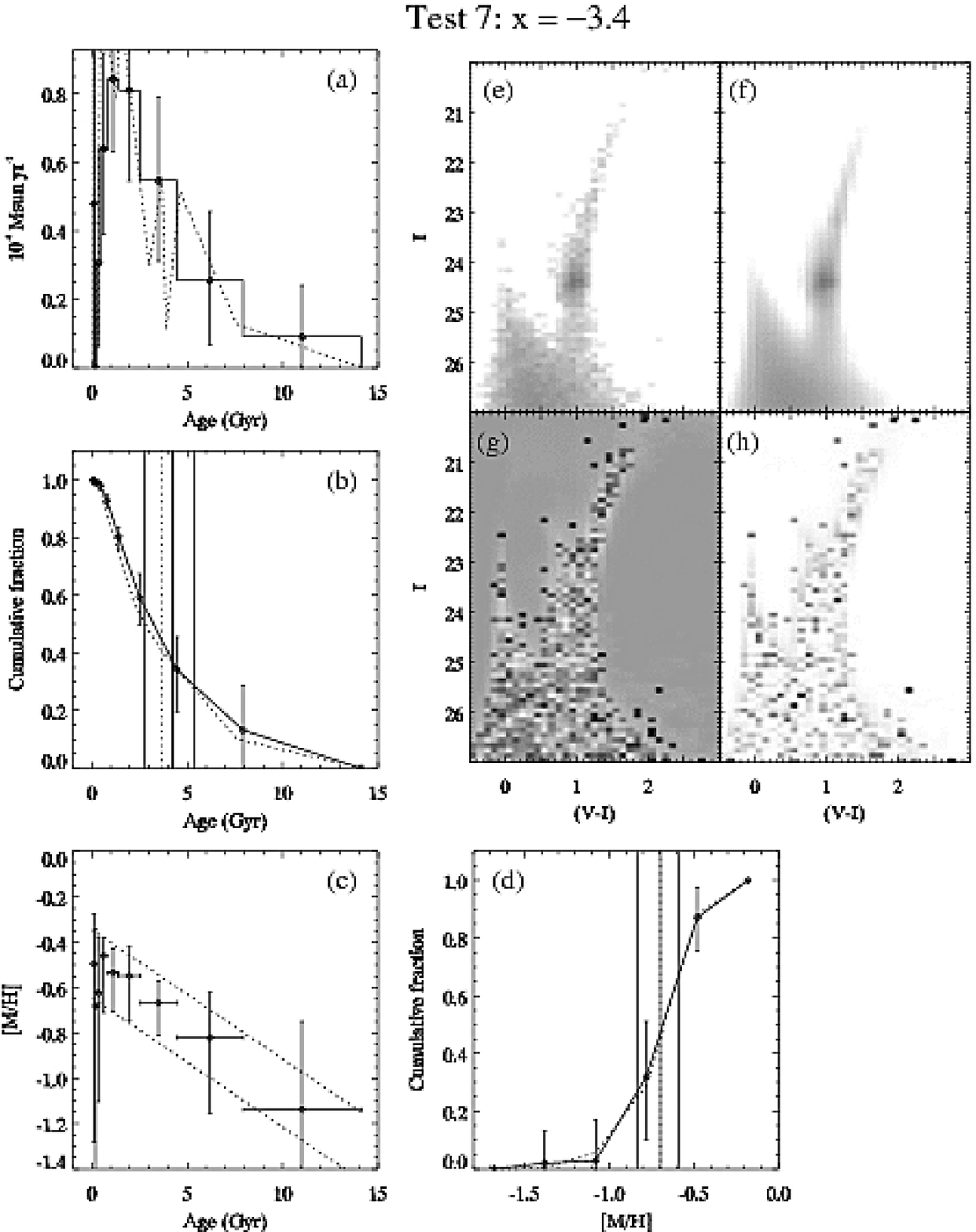}
\caption{Same as Fig.\ \ref{fig:test1}.}
\label{fig:test7}
\end{figure*}

\begin{figure*}
\epsscale{0.95}
\plotone{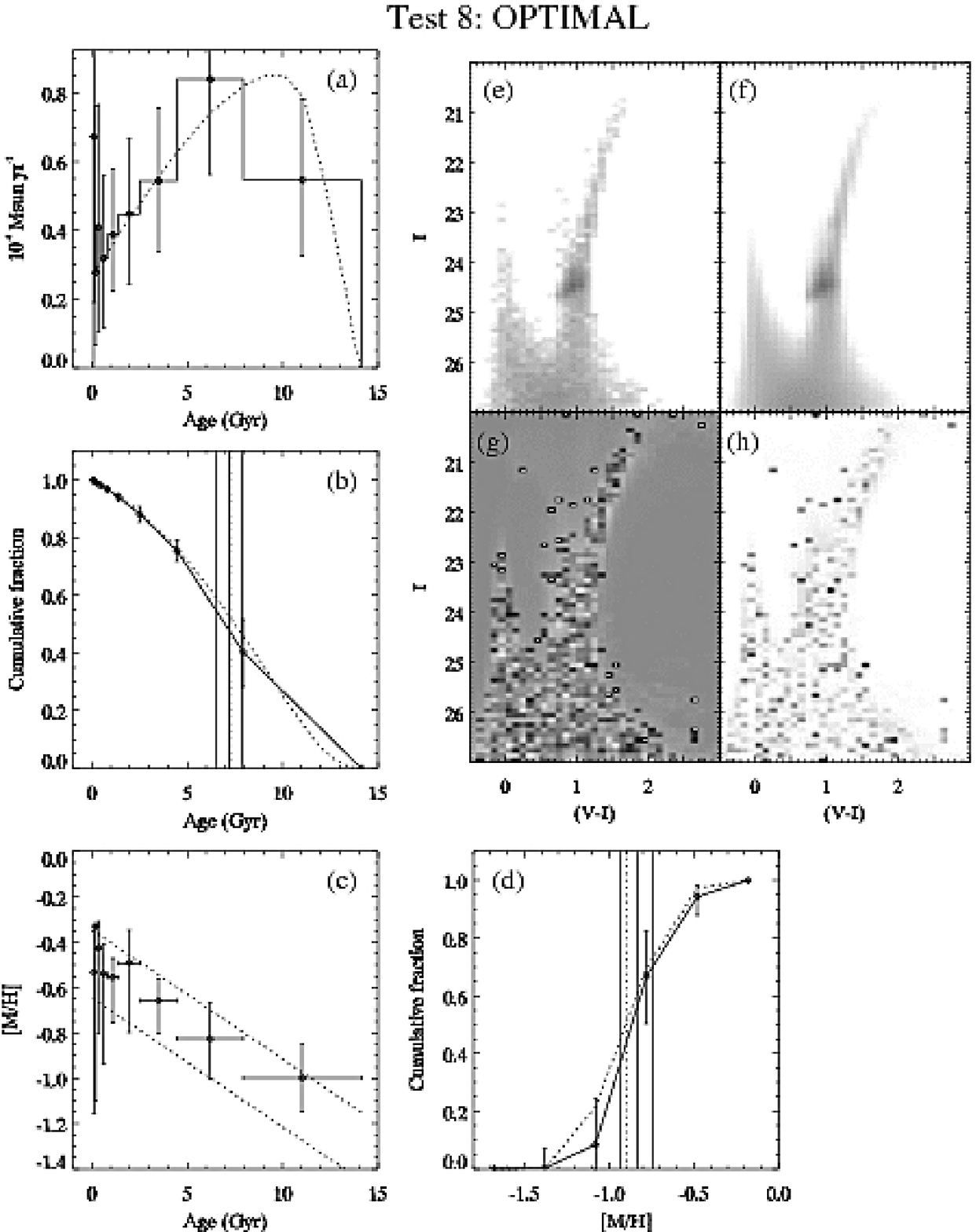}
\caption{Same as Fig.\ \ref{fig:test1}.}
\label{fig:test8}
\end{figure*}

\begin{figure*}
\epsscale{0.95}
\plotone{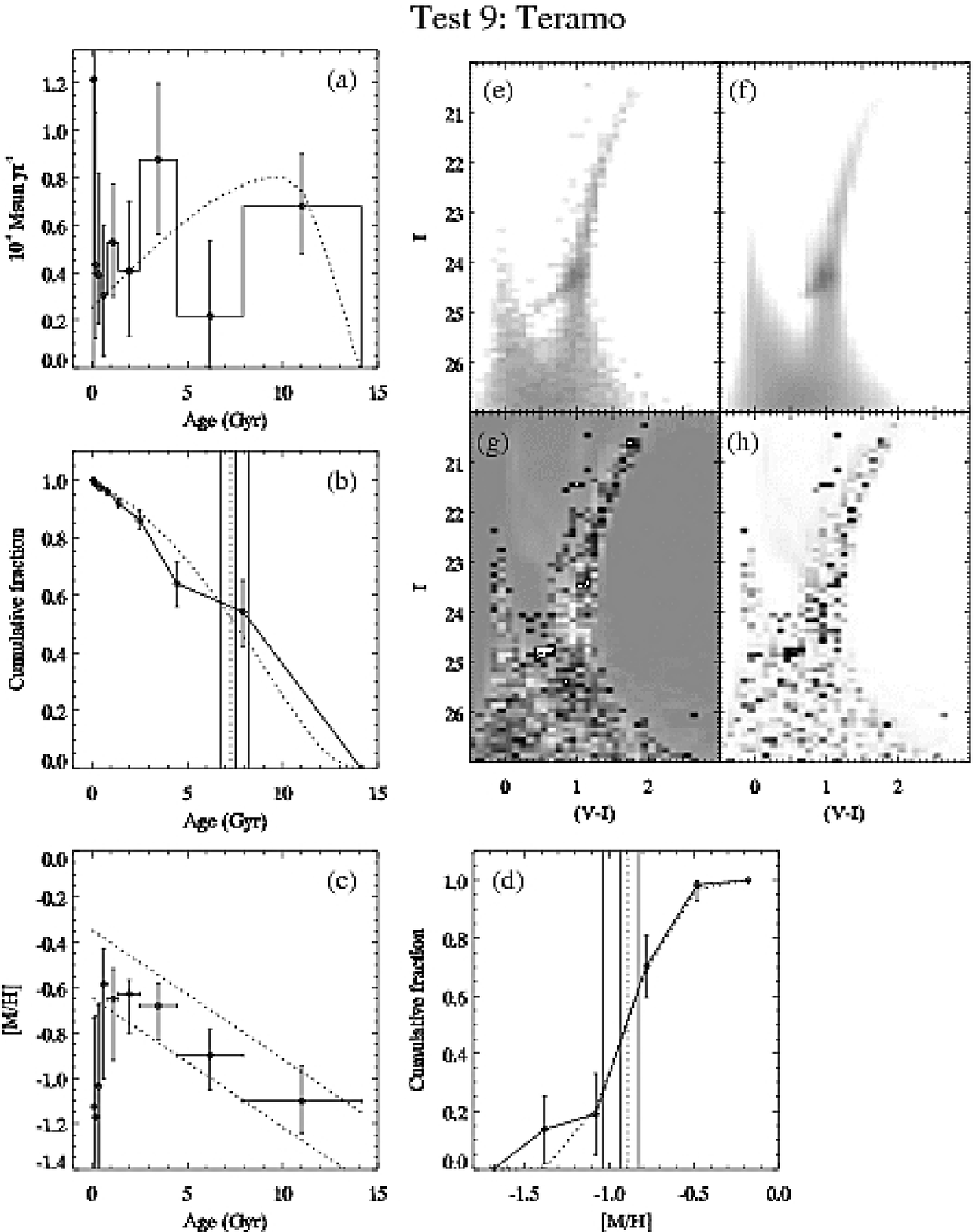}
\caption{Same as Fig.\ \ref{fig:test1}.}
\label{fig:test9}
\end{figure*}

\begin{figure*}
\epsscale{0.95}
\plotone{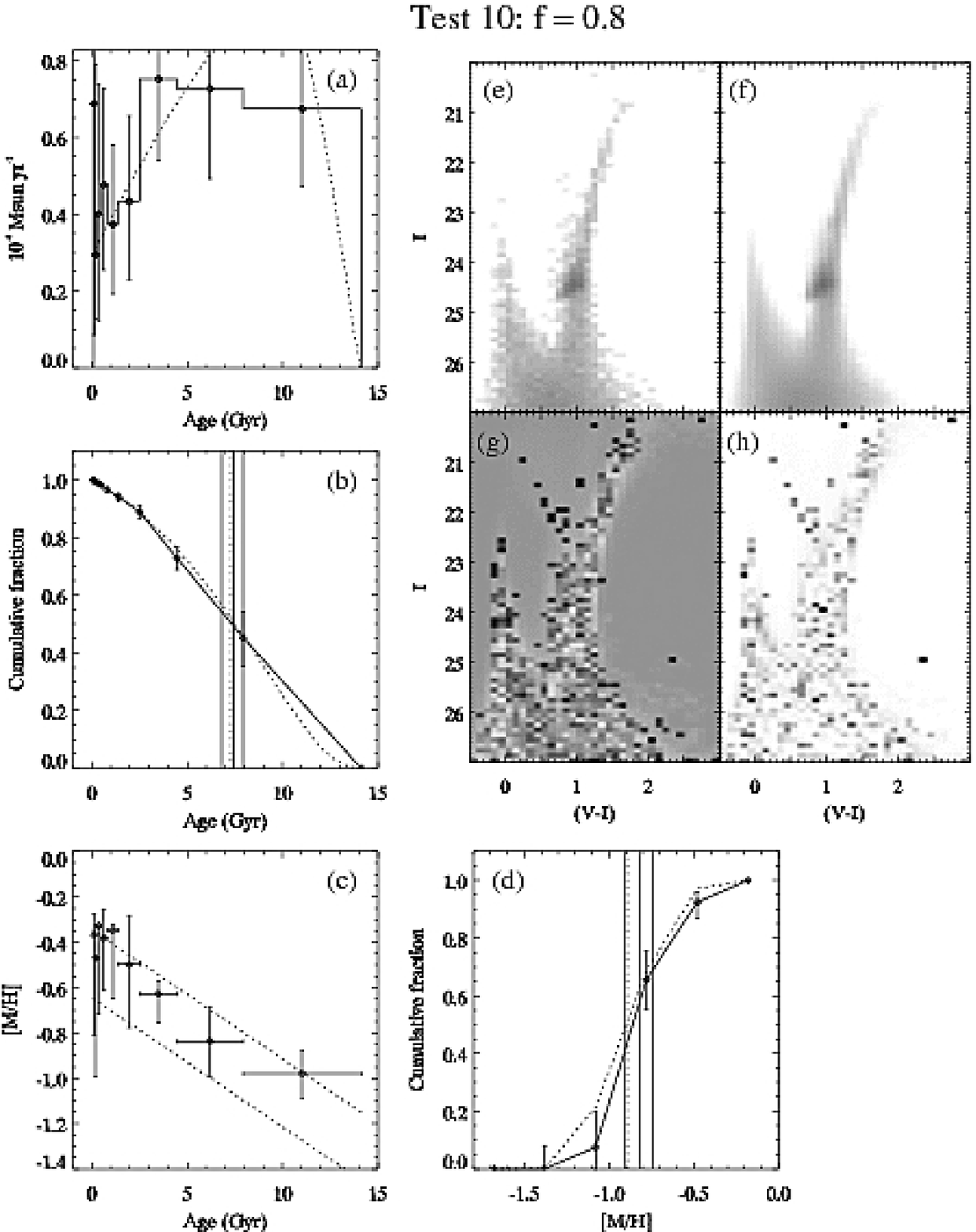}
\caption{Same as Fig.\ \ref{fig:test1}.}
\label{fig:test10}
\end{figure*}

\begin{figure*}
\epsscale{0.95}
\plotone{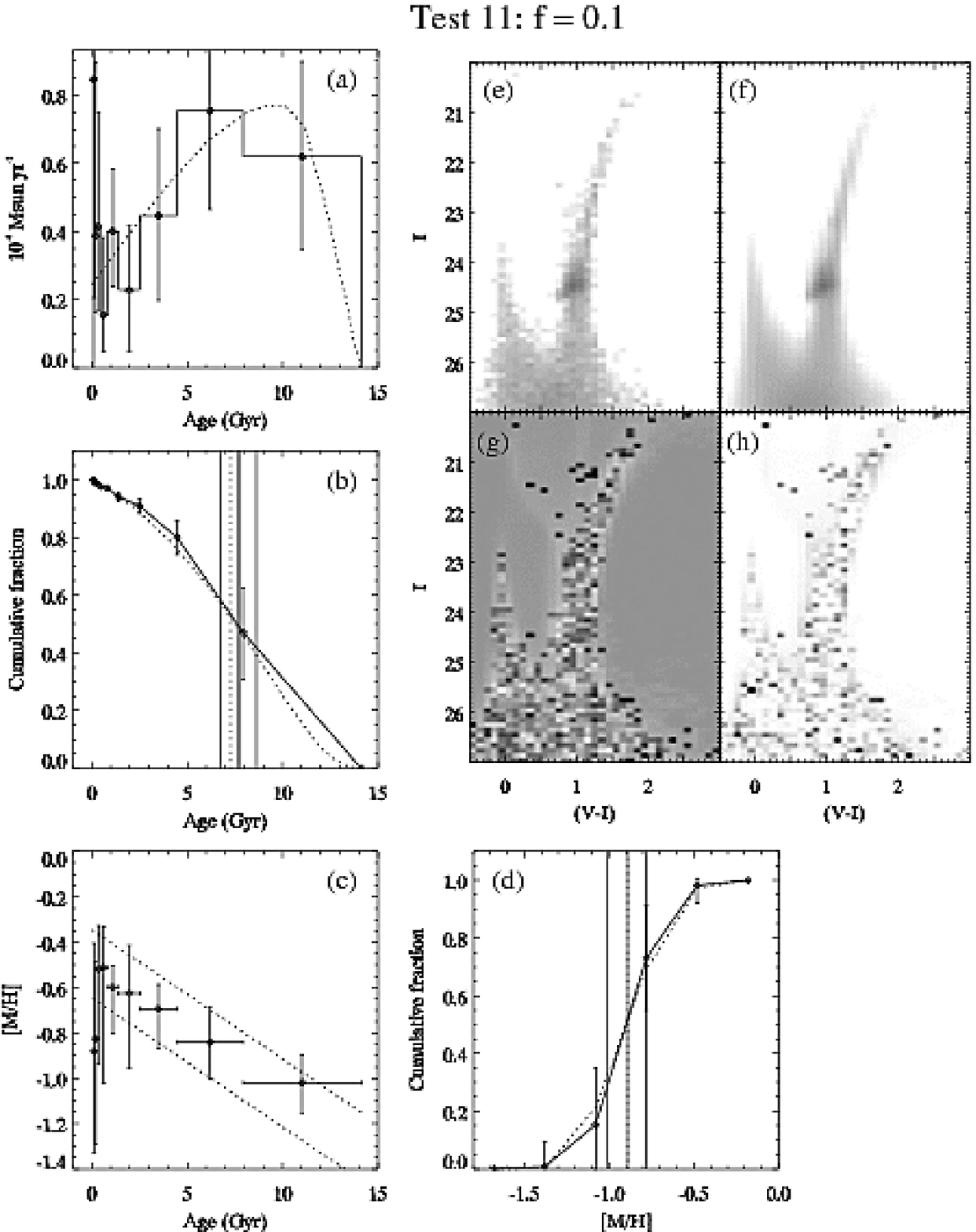}
\caption{Same as Fig.\ \ref{fig:test1}.}
\label{fig:test11}
\end{figure*}

\begin{figure*}
\epsscale{0.95}
\plotone{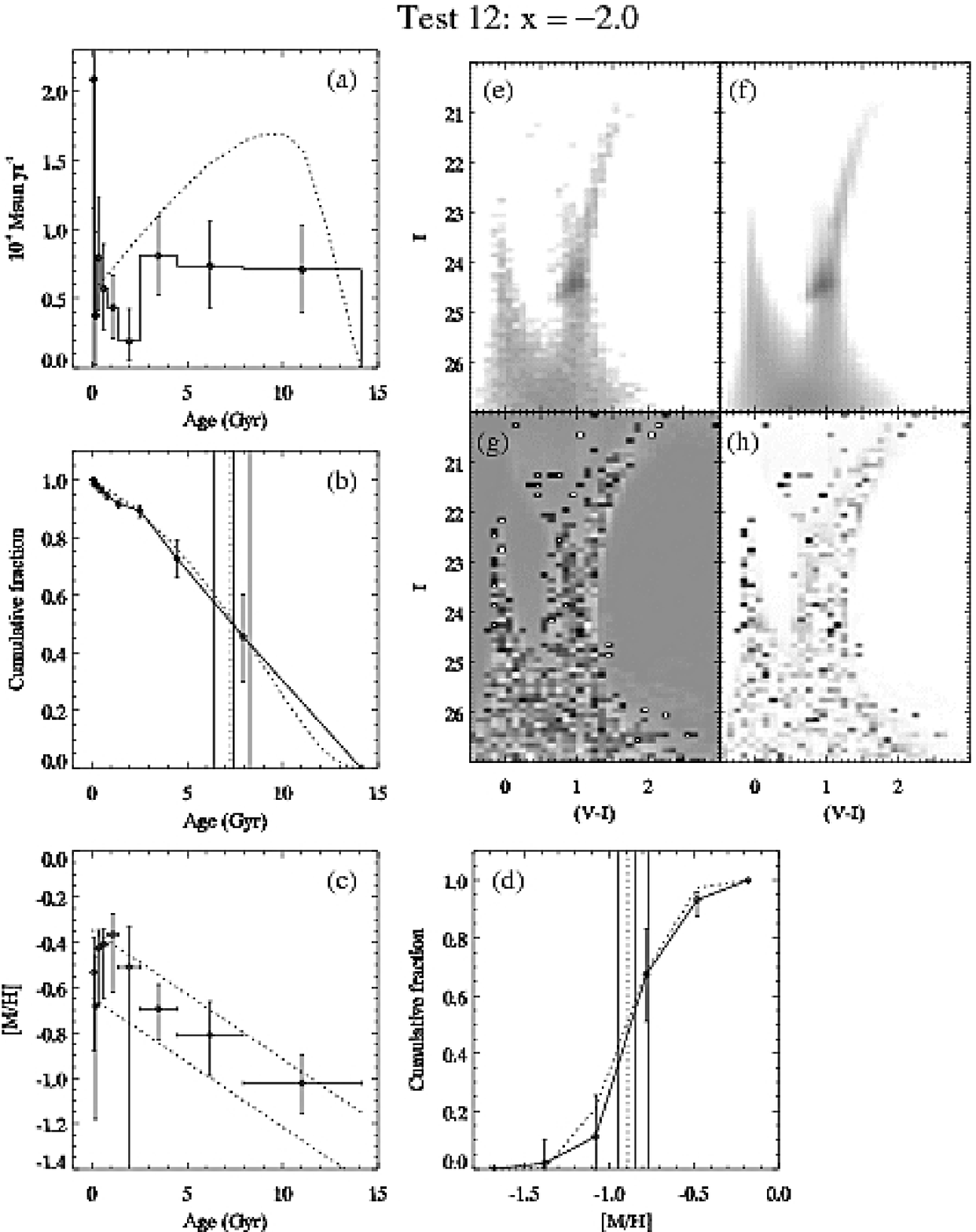}
\caption{Same as Fig.\ \ref{fig:test1}.}
\label{fig:test12}
\end{figure*}

\begin{figure*}
\epsscale{0.95}
\plotone{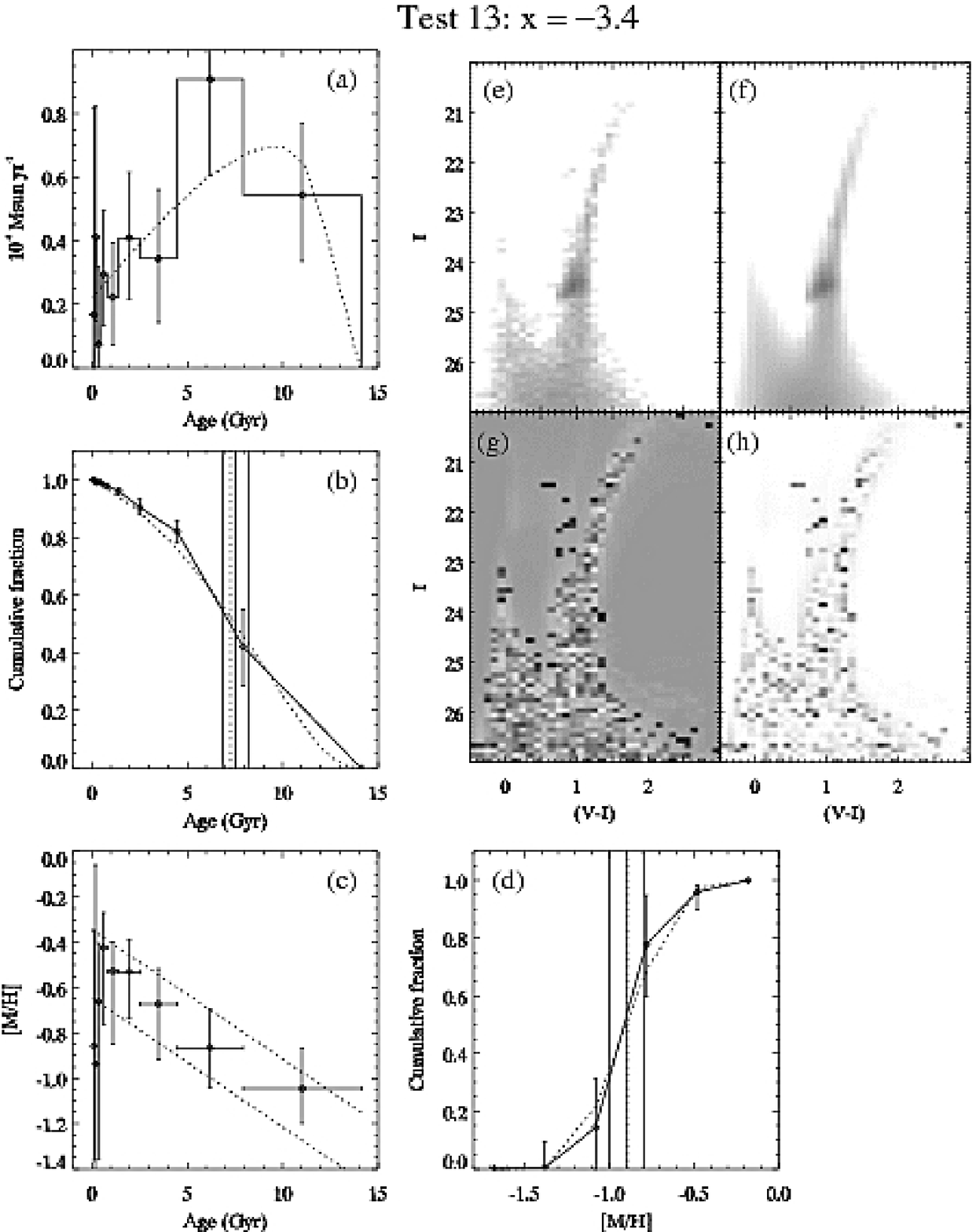}
\caption{Same as Fig.\ \ref{fig:test1}.}
\label{fig:test13}
\end{figure*}

\begin{figure*}
\epsscale{0.95}
\plotone{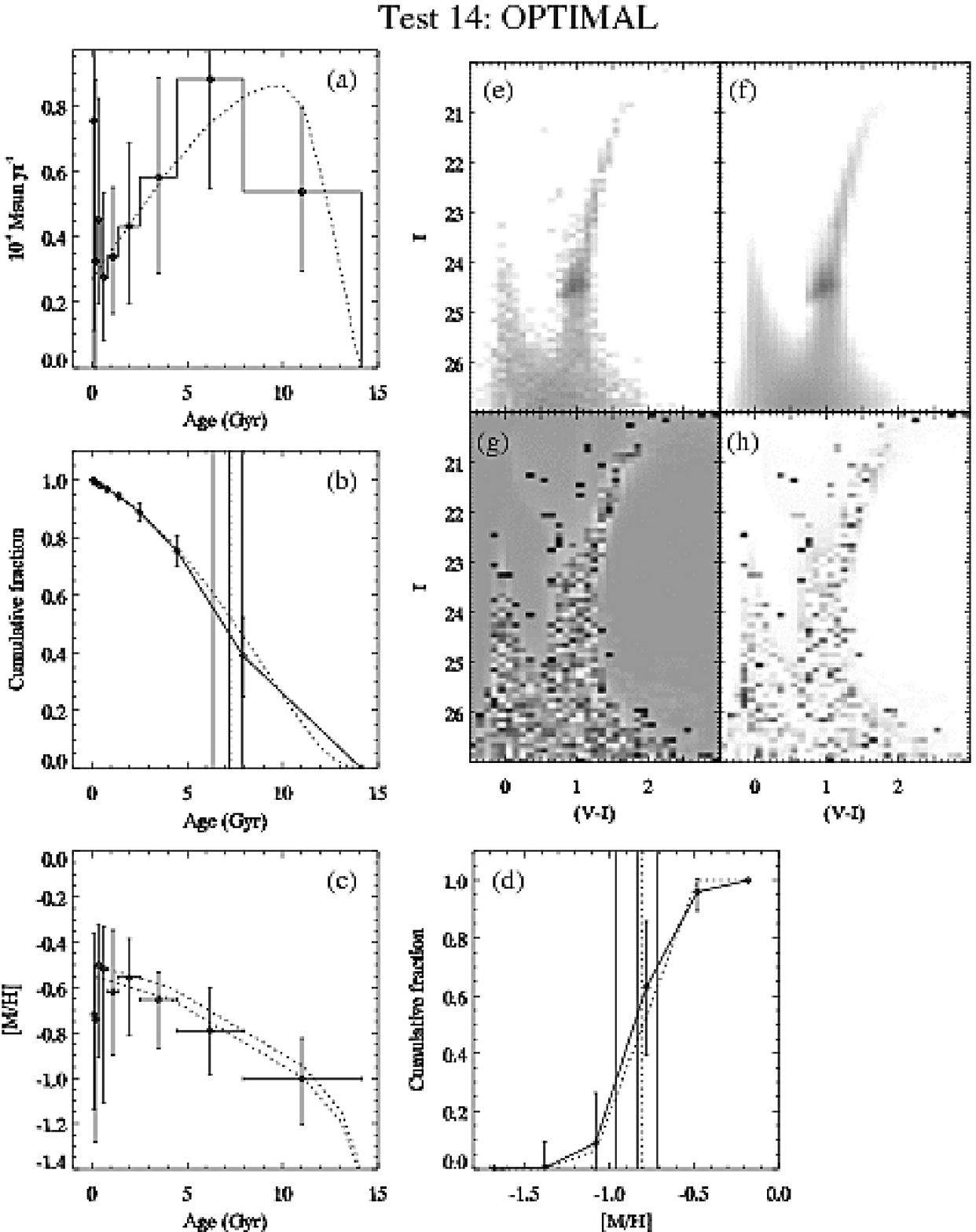}
\caption{Same as Fig.\ \ref{fig:test1}.}
\label{fig:test14}
\end{figure*}

\begin{figure*}
\epsscale{0.95}
\plotone{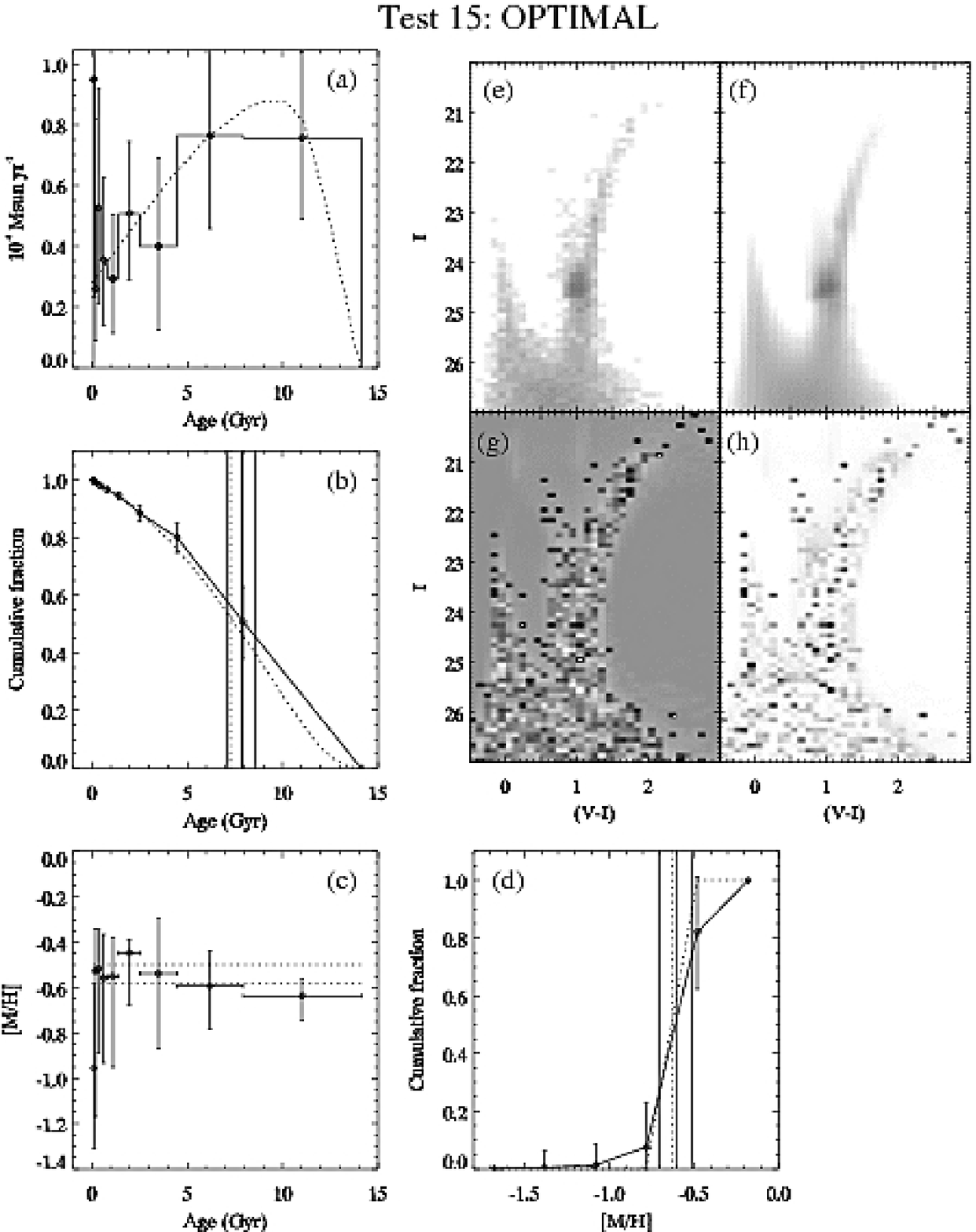}
\caption{Same as Fig.\ \ref{fig:test1}.}
\label{fig:test15}
\end{figure*}

\begin{figure*}
\epsscale{0.95}
\plotone{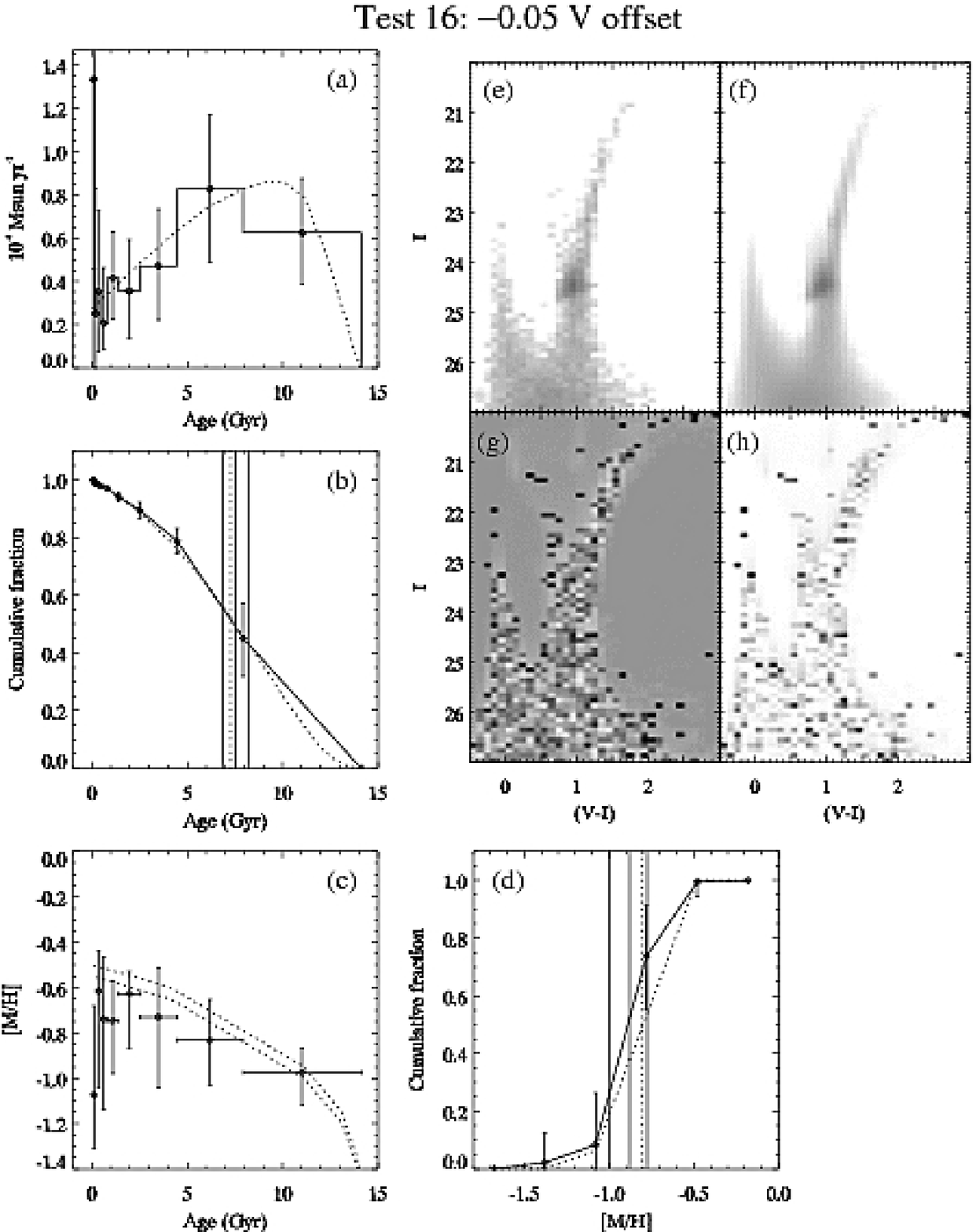}
\caption{Same as Fig.\ \ref{fig:test1}.}
\label{fig:test16}
\end{figure*}

Allowing age and metallicity to vary as free 
parameters and assuming SF began $\sim 14$ Gyr ago, 
we find that, in A1, the
mean SFR $80 - 800$ Myr ago was $\sim 30\%$ as high as
the lifetime-averaged SFR, in A2 it was 
$\sim 10\%$ as high, and in A3 $\sim 5\%$ as high.
Averaging the results from the Padova and Teramo tracks,
the fraction of stars formed by 4.5 Gyr ago
increases from $\sim 65\%$ in A1 to $\sim 80\%$ in A3.
The mean age of all stars and stellar remnants 
increases from $\sim 6$ Gyr to $\sim 8$ Gyr and
the mean global metallicity decreases from $\sim -0.7$ to $\sim -0.9$.  
The random errors on these age and metallicity estimates are 
$\sim 10\%$, 1.0 Gyr, and 0.1 dex.  By comparing the results of
the two sets of stellar tracks for the real data and for
test populations with known SFH we have estimated the 
systematic errors to be $15\%$, 1.0 Gyr, and 0.2 dex.
These do not include uncertainties in the bolometric 
corrections or $\alpha$-element abundances which undoubtedly
deserve future study.

Simple linear least-squares fits to the mean 
ages and metallicities of all three fields
yield an age gradient of 
$\rm 0.58 \pm 0.27(rand) \pm 0.15(sys)\ Gyr\ kpc^{-1}$
and a metallicity gradient of 
$\rm -0.06 \pm 0.05(rand) \pm 0.03(sys)\ dex\ kpc^{-1}$.
This metallicity gradient is roughly consistent with that
found in Paper II modulo an offset of $\sim 0.4$ dex.
Half of this offset is due to the younger mean age
of M33 compared to the GGCs and half is due to the
lower $\alpha$-element abundance ([$\alpha$/Fe] = 0) 
of the stellar tracks (see also Salaris \& Girardi 2005).
We caution that the age and metallicity
gradients do not necessarily continue into the inner disk.
However, the stellar $M/L_V$ implied by our results
is consistent with extrapolation of the independent estimates 
of Ciardullo et al.\ (2004) for regions interior to ours 
and together they imply that $M/L_V$ increases linearly
over $\sim 6$ visual scale lengths, 
from $\sim 0.3$ at $R_{dp} = 1$ kpc 
to $\sim 2.0$ at $R_{dp} = 13$ kpc.

In Paper II we found that the stellar scale length increases 
with age in a roughly power-law fashion reminiscent
of what has been observed in the {\it vertical} direction
in six low-mass spirals (Seth et al.\ 2005).
This behavior could be caused by the orbital diffusion
of stars as they age.
Therefore, the SFH we have derived in the present study could reflect a 
superposition of star formation and later dynamical
processes which act to redistribute stars in the disk.
Any similar analyses carried out on other stellar
populations, especially those of disk galaxies, could
face similar uncertainties.

\section{Appendix}
\label{sec:app}

Here we provide a demonstration of the synthetic CMD-fitting
method used in this paper by fitting many test populations
with known SFHs.
Each test population was generated in IAC-STAR
with $q = 0.60, (m-M)_0 = 24.68$, and $A_V = 0.18$ and the
SFR was normalized to produce $\approx 10,000$ observed stars.
We varied the binary fraction ($f$), the slope of the
IMF for masses above $1.0\ M_{\sun}$ ($x$), and the stellar tracks
used to make the test populations.
In all cases, the distance and extinction were fitted simultaneously
with the SFH in the same manner as for the real data.

The results are shown in Figures $\ref{fig:test1} - \ref{fig:test16}$ 
where the figure
titles tell the true value of the varied parameter.  ``Optimal'' tests
have the parameters held at their true values.
All test populations were generated with the Padova tracks except for those
titled ``Teramo'' which were generated with the Teramo tracks.
The Padova synthetic CMDs were used to fit every test population.
The panels in these 
figures are the same as in Fig.\ \ref{fig:A1_padova_bad} but we also show the 
true SFH, age-CDF, AMR, and Z-CDF as dotted lines.
The vertical dotted lines represent the true mean age and metallicity
of all stars ever formed.
Lastly, Table \ref{tab:tests1} lists the fit quality, 
distance, and extinction for each test.

In general, the agreement between the recovered SFH, distance, 
and extinction and their true values is good.
The error bars are realistic
indicators of the typical deviations.  
Even when the true binary fraction is as low as 0.1
or as high as 0.8 the solution is quite accurate.
Errors in the high-mass IMF slope can cause
normalization errors in the recovered differential SFRs
yet the age-CDF, AMR, and Z-CDF are recovered
accurately and there are no large residuals between
the model and data CMDs.  
Therefore, the method
is stable against reasonable errors in the binary fraction
and IMF given the depth of our photometry.

Due to the relatively large age bins employed in the
present analysis the peak in the recovered SFH can 
be significantly different
from the peak in the true SFH.  For example, even in 
the best-case scenario where all input parameters are
correct, Test 8 shows that one could mistakenly conclude
a peak in the SFR at ages $\sim 4.5 - 8.0$ Gyr when in
fact the peak is at $9 - 10$ Gyr.  
Similarly, the large age bins limit the conclusions that can
be made regarding the ``burstiness'' of the SFH.
Any burst of shorter duration than the corresponding age bin
will be distributed throughout the entire bin. 
Even in the optimal cases adjacent bins
can have the same SFR or metallicity leading one to
think that these quantities are constant when they 
are truly changing.  Therefore, caution 
is required when interpreting bin-to-bin variations and 
more weight must be given to the net change 
over several bins.  
However, the age-CDF is
recovered quite accurately and is relatively insensitive
to errors in $f$ or $x$.  Hence, {\it conclusions based
on the age-CDF are in general more robust than those based
on the differential SFH} (see also Holtzman 2001).

Since the age-sensitive sub-giant branch for ages 
$\gtrsim 5$ Gyr lies fainter than $I = 27$, there 
is not as much information available to distinguish
between the oldest two age bins.  Therefore, these
bins are more susceptible to the age-metallicity
degeneracy and they should be
considered with caution.  The tests show that the largest deviations
often occur in these bins but they are typically
within the error bars.

Another important fact is that the youngest three
points in the SFH and AMR can show significant errors even
when the entire age-CDF and Z-CDF are reasonably recovered.  This
behavior arises from small number statistics in the
CMD regions occupied by stars with the youngest ages.
The metallicity has a small effect on
the position of the bright MS and a much larger effect on
the red supergiant and blue helium-burning phases 
(e.g., Dolphin et al.\ 2003)
but if there are few stars populating the latter then
the metallicity at ages $< 1$ Gyr is hard to constrain.

The largest inaccuracies in the solutions come from
errors in the stellar tracks.  
Tests 3 and 9 show that imperfections in the tracks
can cause deviations larger than the $1\sigma$ errors
although nearly all the deviations are $< 2\sigma$.
The CMD regions that are fit poorly may only contain a
small range of ages or metallicities, thus affecting
a small portion of the recovered SFH.  The strong
correlations between adjacent age/metallicity bins, though, may cause 
errors in one bin to leak into other bins.
More importantly,
there are multiple CMD regions with the same ages and metallicities
so if one region is fit poorly then the others can still drive the
solution toward a good fit (provided there are no errors
in those other regions).  
The stellar tracks are more accurate at some ages and 
metallicities than others.  Therefore, the accuracy of the 
recovered SFH depends on the true SFH itself.  

We investigated several other CMD binning schemes
for Tests 3 and 9 including 0.25 mag square bins 
and rectangular bins longer
in the color dimension.  We also 
tried masking out the RC region from the fits to see
if that was the main source of error.
In all cases the solutions were somewhat less
accurate than in the original binning scheme.
This agrees with the findings of Dolphin (2002) that
increasing the CMD bin size decreases sensitivity and 
that the RC can contain vital age
and metallicity information even when it is not
perfectly modeled.

Finally, in the last test we have applied an offset of
$-0.05$ mag to the V-band of the simulated data stars
with $\rm [M/H] > -0.8$.  Such a metallicity-dependent offset
could arise from an imperfect transformation to the
ground-based photometric system.  The solution
is almost unaffected.  All quantities are recovered
accurately.  The only systematic error occurs in the AMR
where the metallicitiy is underestimated by $\sim 0.1$ dex
for $\rm [M/H] > -0.8$.  Nevertheless, this difference 
is still within the $1\sigma$ errors of the solution.

These tests show that the 
method can reliably extract useful information 
such as the age and metallicity distributions of
all stars ever formed.  This holds even when the
binary fraction and high-mass IMF slope are 
reasonably different from the values we have assumed.  
Errors in the tracks themselves
make the largest contribution to our systematic errors
which we can quantify by comparing the results (for the
real data and test data) obtained with 
the Padova and Teramo tracks.
We estimate conservative systematic
uncertainties of $\pm 15\%$ in the age-CDF, $\pm 1.0$ Gyr in the mean
age, and $\pm 0.2$ dex in the AMR and mean metallicity.  These
estimates do not include variations in the $\alpha$-element
abundances or errors in the bolometric corrections.

\acknowledgements

We warmly thank Jason Harris for invaluable
feedback and help with StarFISH and Antonio Aparicio for 
his generosity and assistance with IAC-STAR.  
We also thank Jon Holtzman for helpful comments on a draft.
This work has made use of the IAC-STAR Synthetic CMD computation code. 
IAC-STAR is suported and maintained by the computer division of the 
Instituto de Astrof\'{i}sica de Canarias.
This research was supported by NSF CAREER grant AST 00-94048 to A.S.
D.G. gratefully acknowledges support from the Chilean
{\sl Centro de Astrof\'\i sica} FONDAP No.\ 15010003.

\clearpage

\clearpage


\begin{deluxetable*}{rrrrrrrr}
\tablewidth{0pt}
\tablecolumns{8}
\tablecaption{Basic results of SFH solutions \label{tab:sfhresults1}}
\tablehead{ \colhead{Field} & \colhead{$Q$} & \colhead{$\chi_{\nu}^2$} & \colhead{$\nu$} & \colhead{$\overline{(m-M)_0}$} & \colhead{$\sigma$} & \colhead{$\overline{A_V}$} & \colhead{$\sigma$} \\
\colhead{} & \colhead{} & \colhead{} & \colhead{} & \colhead{} & \colhead{} & \colhead{} & \colhead{} }
\startdata
\cutinhead{Padova tracks}
A1  &    6.64  &    1.70  & 1714  &   24.60  &    0.05  &    0.20  &    0.06 \\
A2  &    3.98  &    1.57  & 1730  &   24.63  &    0.07  &    0.16  &    0.06 \\
A3  &    2.81  &    1.66  & 1734  &   24.61  &    0.09  &    0.18  &    0.06 \\
\cutinhead{Teramo tracks}
A1  &    6.02  &    1.68  & 1726  &   24.70  &    0.05  &    0.15  &    0.06 \\
A2  &    3.55  &    1.53  & 1727  &   24.75  &    0.07  &    0.15  &    0.05 \\
A3  &    2.99  &    1.68  & 1733  &   24.71  &    0.10  &    0.17  &    0.06 \\
\enddata
\end{deluxetable*}


\begin{deluxetable*}{rrrrrrrrrr}
\tablewidth{0pt}
\tablecolumns{10}
\tablecaption{Basic results of SFH solutions \label{tab:sfhresults2}}
\tablehead{ \colhead{Field} & \colhead{$\overline{\rm Age}$} & \colhead{$\sigma_{hi}$}  & \colhead{$\sigma_{lo}$} & \colhead{$\overline{\rm [M/H]}$} & \colhead{$\sigma_{hi}$} & \colhead{$\sigma_{lo}$} & \colhead{$\overline{M/L_V}$} & \colhead{$\sigma_{hi}$} & \colhead{$\sigma_{lo}$} \\
\colhead{} & \colhead{(Gyr)} & \colhead{(Gyr)} & \colhead{(Gyr)} & \colhead{} & \colhead{} & \colhead{} & \colhead{} & \colhead{} & \colhead{} }
\startdata
\cutinhead{Padova tracks}
A1  & 6.09  &    0.59  &    0.67  &   -0.77  &    0.11  &    0.12  &    1.56  &    0.16  &    0.16\\
A2  & 6.87  &    0.76  &    0.83  &   -0.78  &    0.13  &    0.15  &    1.82  &    0.23  &    0.22\\
A3  & 7.99  &    0.86  &    0.98  &   -0.93  &    0.19  &    0.16  &    2.04  &    0.29  &    0.23\\
\cutinhead{Teramo tracks}
A1  & 6.50  &    0.46  &    0.51  &   -0.66  &    0.11  &    0.11  &    1.53  &    0.13  &    0.13\\
A2  & 7.00  &    0.68  &    0.75  &   -0.77  &    0.10  &    0.13  &    1.62  &    0.20  &    0.19\\
A3  & 8.09  &    0.97  &    1.24  &   -0.88  &    0.18  &    0.18  &    1.83  &    0.31  &    0.29\\

\enddata
\end{deluxetable*}


\tabletypesize{\scriptsize}
\begin{deluxetable*}{rrrrrrrrrr}
\tablecolumns{10}
\tablewidth{0pt}
\tablecaption{SFH of M33's Outer Regions \label{tab:sfh}}

\tablehead{
  \colhead{Age Range} & \multicolumn{3}{c}{A1} & \multicolumn{3}{c}{A2} & \multicolumn{3}{c}{A3} \\ \cline{2-4} \cline{5-7} \cline{8-10} \colhead{log(yr)} & 
      \colhead{$SFR$} & \colhead{$\sigma_{hi}$} & \colhead{$\sigma_{lo}$} &
      \colhead{$SFR$} & \colhead{$\sigma_{hi}$} & \colhead{$\sigma_{lo}$} &
      \colhead{$SFR$} & \colhead{$\sigma_{hi}$} & \colhead{$\sigma_{lo}$}
}
\startdata

\cutinhead{Padova tracks}
 9.90--10.15  &  0.673  &  0.276  &  0.262  &  0.320  &  0.165  &  0.140  &  0.176  &  0.104  &  0.078  \\
 9.65-- 9.90  &  0.783  &  0.318  &  0.290  &  0.286  &  0.218  &  0.209  &  0.126  &  0.112  &  0.097  \\
 9.40-- 9.65  &  2.601  &  0.345  &  0.336  &  0.810  &  0.327  &  0.320  &  0.229  &  0.120  &  0.102  \\
 9.15-- 9.40  &  0.950  &  0.322  &  0.319  &  0.241  &  0.172  &  0.154  &  0.037  &  0.097  &  0.037  \\
 8.90-- 9.15  &  0.680  &  0.261  &  0.235  &  0.192  &  0.133  &  0.099  &  0.044  &  0.082  &  0.044  \\
 8.65-- 8.90  &  0.220  &  0.283  &  0.220  &  0.010  &  0.109  &  0.010  &  0.014  &  0.082  &  0.013  \\
 8.40-- 8.65  &  0.175  &  0.437  &  0.171  &  0.058  &  0.154  &  0.033  &  0.001  &  0.101  &  0.001  \\
 8.15-- 8.40  &  0.371  &  0.609  &  0.318  &  0.013  &  0.228  &  0.013  &  0.001  &  0.162  &  0.001  \\
 7.90-- 8.15  &  0.978  &  1.012  &  0.611  &  0.047  &  0.378  &  0.047  &  0.034  &  0.279  &  0.034  \\
\cutinhead{Teramo tracks}
 9.90--10.15  &  0.606  &  0.252  &  0.229  &  0.272  &  0.161  &  0.134  &  0.160  &  0.109  &  0.091  \\
 9.65-- 9.90  &  1.877  &  0.284  &  0.271  &  0.633  &  0.159  &  0.146  &  0.231  &  0.107  &  0.094  \\
 9.40-- 9.65  &  1.756  &  0.325  &  0.313  &  0.550  &  0.231  &  0.224  &  0.117  &  0.112  &  0.098  \\
 9.15-- 9.40  &  0.495  &  0.247  &  0.223  &  0.178  &  0.114  &  0.094  &  0.041  &  0.072  &  0.028  \\
 8.90-- 9.15  &  0.241  &  0.275  &  0.235  &  0.022  &  0.112  &  0.020  &  0.003  &  0.077  &  0.003  \\
 8.65-- 8.90  &  0.367  &  0.335  &  0.277  &  0.047  &  0.131  &  0.047  &  0.007  &  0.093  &  0.006  \\
 8.40-- 8.65  &  0.191  &  0.471  &  0.164  &  0.045  &  0.176  &  0.024  &  0.002  &  0.123  &  0.002  \\
 8.15-- 8.40  &  0.115  &  0.655  &  0.115  &  0.001  &  0.245  &  0.001  &  0.001  &  0.189  &  0.001  \\
 7.90-- 8.15  &  1.052  &  1.121  &  0.881  &  0.014  &  0.386  &  0.014  &  0.025  &  0.304  &  0.015  \\

\enddata
\end{deluxetable*}
\tablecomments{The units are $10^{-4}\ M_{\sun}\ \rm yr^{-1}$.}

\clearpage

\tabletypesize{\scriptsize}
\begin{deluxetable*}{rrrrrrrrrr}
\tablecolumns{10}
\tablewidth{0pt}
\tablecaption{Age-CDF of M33's Outer Regions \label{tab:age-cdf}}

\tablehead{
  \colhead{Age Range} & \multicolumn{3}{c}{A1} & \multicolumn{3}{c}{A2} & \multicolumn{3}{c}{A3} \\ \cline{2-4} \cline{5-7} \cline{8-10} \colhead{log(yr)} & 
     \colhead{$M/M_{tot}$} & \colhead{$\sigma_{hi}$} & \colhead{$\sigma_{lo}$} & 
     \colhead{$M/M_{tot}$} & \colhead{$\sigma_{hi}$} & \colhead{$\sigma_{lo}$} & 
     \colhead{$M/M_{tot}$} & \colhead{$\sigma_{hi}$} & \colhead{$\sigma_{lo}$}
}
\startdata

\cutinhead{Padova tracks}
 9.90--10.15  &  0.304  &  0.082  &  0.093  &  0.397  &  0.096  &  0.108  &  0.529  &  0.131  &  0.149  \\
 9.65-- 9.90  &  0.504  &  0.060  &  0.068  &  0.595  &  0.128  &  0.131  &  0.744  &  0.091  &  0.092  \\
 9.40-- 9.65  &  0.877  &  0.024  &  0.023  &  0.919  &  0.028  &  0.030  &  0.963  &  0.024  &  0.048  \\
 9.15-- 9.40  &  0.954  &  0.011  &  0.011  &  0.972  &  0.013  &  0.016  &  0.983  &  0.013  &  0.024  \\
 8.90-- 9.15  &  0.984  &  0.006  &  0.007  &  0.996  &  0.001  &  0.007  &  0.996  &  0.002  &  0.014  \\
 8.65-- 8.90  &  0.990  &  0.003  &  0.006  &  0.997  &  0.001  &  0.006  &  0.999  &  0.001  &  0.010  \\
 8.40-- 8.65  &  0.993  &  0.003  &  0.005  &  0.999  &  0.001  &  0.005  &  0.999  &  0.001  &  0.009  \\
 8.15-- 8.40  &  0.996  &  0.003  &  0.005  &  0.999  &  0.001  &  0.004  &  0.999  &  0.001  &  0.008  \\
 7.90-- 8.15  &  1.000  &  0.000  &  0.000  &  1.000  &  0.000  &  0.000  &  1.000  &  0.000  &  0.000  \\
\cutinhead{Teramo tracks}
 9.90--10.15  &  0.256  &  0.073  &  0.081  &  0.324  &  0.107  &  0.121  &  0.476  &  0.167  &  0.215  \\
 9.65-- 9.90  &  0.702  &  0.040  &  0.042  &  0.747  &  0.079  &  0.079  &  0.864  &  0.087  &  0.086  \\
 9.40-- 9.65  &  0.936  &  0.016  &  0.018  &  0.955  &  0.019  &  0.022  &  0.975  &  0.014  &  0.036  \\
 9.15-- 9.40  &  0.973  &  0.010  &  0.011  &  0.992  &  0.004  &  0.013  &  0.997  &  0.002  &  0.022  \\
 8.90-- 9.15  &  0.983  &  0.006  &  0.008  &  0.995  &  0.004  &  0.009  &  0.998  &  0.001  &  0.015  \\
 8.65-- 8.90  &  0.992  &  0.004  &  0.006  &  0.998  &  0.001  &  0.007  &  0.999  &  0.001  &  0.011  \\
 8.40-- 8.65  &  0.995  &  0.004  &  0.005  &  1.000  &  0.000  &  0.005  &  0.999  &  0.000  &  0.010  \\
 8.15-- 8.40  &  0.996  &  0.004  &  0.005  &  1.000  &  0.000  &  0.005  &  0.999  &  0.000  &  0.009  \\
 7.90-- 8.15  &  1.000  &  0.000  &  0.000  &  1.000  &  0.000  &  0.000  &  1.000  &  0.000  &  0.000  \\

\enddata
\end{deluxetable*}


\tabletypesize{\scriptsize}
\begin{deluxetable*}{rrrrrrrrrr}
\tablecolumns{10}
\tablewidth{0pt}
\tablecaption{AMR of M33's Outer Regions \label{tab:amr}}

\tablehead{
  \colhead{Age Range} & \multicolumn{3}{c}{A1} & \multicolumn{3}{c}{A2} & \multicolumn{3}{c}{A3} \\ \cline{2-4} \cline{5-7} \cline{8-10} \colhead{log(yr)} & 
      \colhead{$\rm [M/H]$} & \colhead{$\sigma_{hi}$} & \colhead{$\sigma_{lo}$} &
      \colhead{$\rm [M/H]$} & \colhead{$\sigma_{hi}$} & \colhead{$\sigma_{lo}$} &
      \colhead{$\rm [M/H]$} & \colhead{$\sigma_{hi}$} & \colhead{$\sigma_{lo}$}
}
\startdata

\cutinhead{Padova tracks}
 9.90--10.15  & -0.980  &  0.197  &  0.197  & -0.902  &  0.174  &  0.208  & -1.105  &  0.295  &  0.196  \\
 9.65-- 9.90  & -0.844  &  0.162  &  0.185  & -0.811  &  0.285  &  0.310  & -0.893  &  0.316  &  0.298  \\
 9.40-- 9.65  & -0.653  &  0.106  &  0.122  & -0.589  &  0.144  &  0.219  & -0.613  &  0.168  &  0.272  \\
 9.15-- 9.40  & -0.528  &  0.171  &  0.232  & -0.713  &  0.333  &  0.283  & -0.887  &  0.339  &  0.622  \\
 8.90-- 9.15  & -0.513  &  0.064  &  0.169  & -0.575  &  0.138  &  0.365  & -0.448  &  0.228  &  0.514  \\
 8.65-- 8.90  & -0.607  &  0.156  &  0.354  & -0.763  &  0.371  &  0.809  & -0.567  &  0.282  &  0.797  \\
 8.40-- 8.65  & -0.449  &  0.274  &  0.723  & -0.727  &  0.373  &  0.678  & -0.641  &  0.330  &  0.962  \\
 8.15-- 8.40  & -0.683  &  0.435  &  0.716  & -1.165  &  0.951  &  0.352  & -0.764  &  0.668  &  0.612  \\
 7.90-- 8.15  & -0.611  &  0.196  &  0.461  & -1.331  &  0.893  &  0.256  & -1.273  &  0.803  &  0.259  \\
\cutinhead{Teramo tracks}
 9.90--10.15  & -0.761  &  0.132  &  0.158  & -0.864  &  0.144  &  0.211  & -0.970  &  0.319  &  0.220  \\
 9.65-- 9.90  & -0.767  &  0.184  &  0.185  & -0.849  &  0.178  &  0.203  & -0.887  &  0.269  &  0.269  \\
 9.40-- 9.65  & -0.437  &  0.072  &  0.098  & -0.523  &  0.135  &  0.215  & -0.656  &  0.254  &  0.342  \\
 9.15-- 9.40  & -0.351  &  0.041  &  0.224  & -0.483  &  0.103  &  0.296  & -0.439  &  0.189  &  0.514  \\
 8.90-- 9.15  & -0.328  &  0.001  &  0.456  & -0.917  &  0.565  &  0.596  & -0.939  &  0.712  &  0.652  \\
 8.65-- 8.90  & -0.448  &  0.180  &  0.334  & -0.666  &  0.239  &  0.808  & -0.680  &  0.897  &  0.378  \\
 8.40-- 8.65  & -0.425  &  0.321  &  0.919  & -0.584  &  0.233  &  1.028  & -0.655  &  0.228  &  1.168  \\
 8.15-- 8.40  & -0.425  &  0.150  &  0.838  & -1.123  &  1.035  &  0.329  & -1.015  &  0.911  &  0.424  \\
 7.90-- 8.15  & -0.329  &  0.002  &  0.475  & -0.951  &  0.473  &  0.882  & -1.006  &  0.482  &  0.804  \\

\enddata
\end{deluxetable*}

\clearpage

\tabletypesize{\scriptsize}
\begin{deluxetable*}{rrrrrrrrrr}
\tablecolumns{10}
\tablewidth{0pt}
\tablecaption{Z-CDF of M33's Outer Regions \label{tab:z-cdf}}

\tablehead{
  \colhead{[M/H]} & \multicolumn{3}{c}{A1} & \multicolumn{3}{c}{A2} & \multicolumn{3}{c}{A3} \\ \cline{2-4} \cline{5-7} \cline{8-10} \colhead{} & 
      \colhead{$M/M_{tot}$} & \colhead{$\sigma_{hi}$} & \colhead{$\sigma_{lo}$} &
      \colhead{$M/M_{tot}$} & \colhead{$\sigma_{hi}$} & \colhead{$\sigma_{lo}$} &
      \colhead{$M/M_{tot}$} & \colhead{$\sigma_{hi}$} & \colhead{$\sigma_{lo}$}
}
\startdata

\cutinhead{Padova tracks}
-1.68--(-1.38)  &  0.085  &  0.085  &  0.061  &  0.143  &  0.123  &  0.089  &  0.194  &  0.186  &  0.194  \\
-1.38--(-1.08)  &  0.120  &  0.115  &  0.092  &  0.177  &  0.145  &  0.108  &  0.327  &  0.199  &  0.223  \\
-1.08--(-0.78)  &  0.412  &  0.154  &  0.158  &  0.429  &  0.193  &  0.203  &  0.684  &  0.133  &  0.185  \\
-0.78--(-0.48)  &  0.874  &  0.078  &  0.095  &  0.754  &  0.124  &  0.121  &  0.812  &  0.097  &  0.183  \\
-0.48--(-0.18)  &  1.000  &  0.000  &  0.000  &  1.000  &  0.000  &  0.000  &  1.000  &  0.000  &  0.000  \\
\cutinhead{Teramo tracks}
-1.68--(-1.38)  &  0.032  &  0.059  &  0.032  &  0.085  &  0.120  &  0.061  &  0.104  &  0.203  &  0.087  \\
-1.38--(-1.08)  &  0.037  &  0.083  &  0.037  &  0.169  &  0.143  &  0.095  &  0.329  &  0.244  &  0.279  \\
-1.08--(-0.78)  &  0.360  &  0.210  &  0.213  &  0.503  &  0.153  &  0.133  &  0.592  &  0.167  &  0.221  \\
-0.78--(-0.48)  &  0.682  &  0.089  &  0.096  &  0.717  &  0.092  &  0.108  &  0.832  &  0.101  &  0.180  \\
-0.48--(-0.18)  &  1.000  &  0.000  &  0.000  &  1.000  &  0.000  &  0.000  &  1.000  &  0.000  &  0.000  \\

\enddata
\end{deluxetable*}


\begin{deluxetable*}{rrrccccc}
\tablecaption{Test population results \label{tab:tests1}}
\tablewidth{0pt}
\tablehead{\colhead{Test} &\colhead{$Q$} &\colhead{$\chi^2_{\nu}$} &\colhead{$\nu$} &\colhead{$\overline{(m-M)_0}$} &\colhead{$\sigma$} &\colhead{$\overline{A_V}$} &\colhead{$\sigma$} }
\startdata
1  &   -0.23  &    0.74  & 1728  &   24.69  &    0.07  &    0.19  &    0.05\\
2  &    0.33  &    0.92  & 1728  &   24.67  &    0.07  &    0.19  &    0.06\\
3  &    5.33  &    1.37  & 1727  &   24.56  &    0.09  &    0.25  &    0.03\\
4  &    0.01  &    0.78  & 1725  &   24.64  &    0.07  &    0.18  &    0.06\\
5  &    0.98  &    0.75  & 1733  &   24.70  &    0.05  &    0.22  &    0.04\\
6  &    0.37  &    0.78  & 1726  &   24.66  &    0.07  &    0.18  &    0.06\\
7  &    0.79  &    0.88  & 1732  &   24.70  &    0.05  &    0.15  &    0.06\\
8  &   -0.90  &    1.27  & 1727  &   24.70  &    0.05  &    0.14  &    0.06\\
9  &    8.51  &    1.59  & 1726  &   24.55  &    0.08  &    0.20  &    0.05\\
10  &    3.15  &    1.02  & 1729  &   24.70  &    0.05  &    0.13  &    0.04\\
11  &    0.98  &    0.90  & 1727  &   24.67  &    0.07  &    0.17  &    0.06\\
12  &    3.00  &    1.13  & 1726  &   24.68  &    0.07  &    0.15  &    0.05\\
13  &    0.24  &    0.61  & 1730  &   24.70  &    0.05  &    0.16  &    0.06\\
14  &    2.70  &    1.21  & 1725  &   24.70  &    0.05  &    0.18  &    0.07\\
15  &    1.41  &    1.07  & 1722  &   24.68  &    0.07  &    0.17  &    0.07\\
16  &    1.64  &    1.33  & 1726  &   24.70  &    0.05  &    0.14  &    0.06\\
\enddata
\end{deluxetable*}

\clearpage


\end{document}